\documentclass[a4paper,11pt]{article}

\usepackage{draft}
\usepackage{amsmath}
\usepackage{amssymb}
\usepackage{amsthm}
\usepackage{mathtools}
\usepackage{mathrsfs}
\usepackage{booktabs,array}
\usepackage{tikz}
\usetikzlibrary{arrows.meta,calc}
\usepackage{microtype}
\usepackage{orcidlink}
\usepackage{cite}
\usepackage{xcolor}
\definecolor{twilightlavender}{rgb}{0.54, 0.29, 0.42}
\definecolor{richmaroon}{rgb}{0.69, 0.19, 0.38}
\definecolor{forestgreen(web)}{rgb}{0.13, 0.55, 0.13}
\definecolor{lava}{rgb}{0.81, 0.06, 0.13}
\hypersetup{
  breaklinks,
  colorlinks,
  citecolor=forestgreen(web),
  filecolor=richmaroon,
  linkcolor=lava,
  urlcolor=twilightlavender,
  pdftitle={Carroll limits of bosonic and supersymmetric D-branes in the Polyakov formulation},
  pdfauthor={Limin Zeng},
  pdfsubject={Carroll contraction and auxiliary reduction in the Polyakov formulation},
  pdfkeywords={Carrollian symmetry, D-branes, Born--Infeld theory,
    constrained Hamiltonian systems}
}

\newcommand{\dd}{\mathrm d}
\newcommand{\tr}{\operatorname{tr}}
\newcommand{\diag}{\operatorname{diag}}

\newcommand{\one}{\mathbf 1}
\newcommand{\spat}{\mathrm{sp}}
\newcommand{\M}{\mathrm M}
\newcommand{\E}{\mathrm E}

\numberwithin{equation}{section}
\renewcommand{\theequation}{\arabic{section}.\arabic{equation}}

\title{Carroll limits of bosonic and supersymmetric\\ D-branes in the Polyakov formulation}

\affiliation[a]{
  School of Fundamental Physics and Mathematical Sciences,\\
  Hangzhou Institute for Advanced Study, UCAS, Hangzhou 310024, China
}
\affiliation[b]{
  Institute of Theoretical Physics, Chinese Academy of Sciences,
  Beijing 100190, China
}
\affiliation[c]{
  University of Chinese Academy of Sciences, Beijing 100049, China
}

\author[a,b,c \orcidlink{0009-0003-5703-2687}]{Limin Zeng}
\emailAdd{zenglimin25@mails.ucas.ac.cn}

\abstract{We derive the magnetic and electric Carroll limits of bosonic and supersymmetric D-branes in the Polyakov formulation at fixed positive rescaled tension. Eliminating the nonsymmetric auxiliary tensor recovers the full Dirac--Born--Infeld action. On the regular domain with a positive-definite induced spatial metric and a positive lapse bounded away from zero, the auxiliary constraints remain second class in both limits. Eliminating them before or after contraction gives the same local reduced phase space. The magnetic Hamiltonian retains the full spatial determinant and the electric Hamiltonian retains the electric displacement squared. We compare their dynamics on flat D2 solutions and on curves in a common warped Carroll background. Both limits retain sixteen independent \(\kappa\)-symmetry parameters for flat Type II D2--D7 branes. The electric limit also retains a Neveu--Schwarz flux term in the fermionic constraint and supercharge matrices. Spatial T-duality commutes with both contractions on fixed-radius circle zero modes when the dual transverse canonical pair is rescaled in the electric case. The prescribed local Carroll--Weyl transformation fails to preserve the constraints, and both actions admit exact global dilations.}

\keywords{Carrollian symmetry, D-branes, Born--Infeld theory, constrained Hamiltonian systems}

\begin{document}
\frenchspacing
\maketitle

\section{Introduction}

We work in flat Type II superspace. Its body, obtained by setting the odd coordinates to zero, is the ten-dimensional target spacetime. For a D\(p\)-brane, \(p\) counts spatial worldvolume dimensions.
\[
\dim\mathcal M_{\rm target}=10,\qquad
\dim\Sigma_{p+1}=p+1 .
\]
A D2 has a three-dimensional worldvolume and seven transverse spatial directions. A D3 has a four-dimensional worldvolume and six transverse spatial directions. We contract one target-time direction while keeping the worldvolume dimension fixed. The resulting target metric is degenerate along the time direction, as in Carroll geometry \cite{DuvalGibbonsHorvathyZhang2014}. The spatial embedding can then be time independent even when its momentum evolves under stresses from the Dirac--Born--Infeld (DBI) determinant. The DBI structure arises in the sigma model of an open string with Dirichlet and Neumann boundary conditions \cite{Leigh1989}.

We call the two D-brane contractions magnetic and electric according to which gauge contribution survives in the Hamiltonian constraint. The magnetic branch retains the spatial field strength inside the DBI determinant. The electric branch retains the square of the electric displacement and a spatial volume term. Such a distinction between contractions also arises in Hamiltonian field theory \cite{HenneauxSalgado2021}. At fixed background, the small-speed-of-light expansion of Born--Infeld electrodynamics has an electric leading term, with magnetic terms entering at subleading order \cite{Mehra2024}. On an unstable (non-BPS) D-brane, the stringy Carroll contraction rescales time and one longitudinal spatial coordinate. Keeping the gauge potential unscaled leaves the magnetic field in the spatial determinant. Scaling the potential and its momentum by inverse powers instead retains the electric displacement squared \cite{Kluson2017}. The corresponding non-BPS constraint algebras and the proposed local rescaling of the transverse embedding, gauge field and tachyon have also been studied \cite{ZengPhase2026}.

Auxiliary fields have been used to construct inequivalent target-space Carroll strings from the Polyakov action with a relativistic worldsheet \cite{HarksenHidalgoSybesmaThorlacius2024}. There are also examples in which classically equivalent relativistic string formulations yield Carroll theories with different numbers of degrees of freedom \cite{CasalbuoniDominiciGomis2024}. These results motivate checking whether auxiliary-field elimination commutes with the Carroll limit at the level of constraints and symplectic structure. For this purpose we use the nonsymmetric-tensor Polyakov formulation of Abou Zeid and Hull \cite{AbouZeidHull1997}. At fixed auxiliary tensor the action is linear in the field strength, and eliminating the tensor recovers the full classical DBI determinant. The original tensor becomes singular in the Carroll limit. In rescaled variables, the auxiliary constraints remain second class for bounded data, fixed positive rescaled tension, a uniformly positive-definite spatial metric and a positive lapse bounded away from zero. For each branch, both orders of operations give the same symplectic form and gauge generators, and therefore the same local reduced phase space.

Sections~\ref{sec:parent}--\ref{sec:reduction} derive the two contractions and establish the equivalence of the two orders. Section~\ref{sec:examples} compares their dynamics in target-time gauge. Section~\ref{sec:super} derives the two Type II actions and classifies their fermionic constraints. Section~\ref{sec:spatial-duality} compares the two branches under spatial T-duality, and Sec.~\ref{sec:carroll-weyl} tests local Carroll--Weyl scaling and derives the global dilations.

\section{Hamiltonian formulation of the auxiliary action}
\label{sec:parent}

Consider the bosonic DBI theory in a flat ten-dimensional target with mostly plus signature and \(p>1\) spatial worldvolume dimensions. Worldvolume indices are \(a,b\), with \(i,j\) ranging over the \(p\) spatial directions. Target indices are \(\mu,\nu=0,\ldots,9\), and \(I,J=1,\ldots,9\) label the spatial target directions. A dot denotes \(\partial_0\), and \(\simeq\) denotes equality on the relevant constraint surface. Parentheses and square brackets on indices denote symmetrization and antisymmetrization with unit weight. The spatial block of a worldvolume tensor carries the subscript \({\rm sp}\). We suppress the coordinate measure in integrals of spatial densities. We absorb the constant dilaton into the tension and the Born--Infeld normalization into the gauge potential. We fix the time endpoints and use closed spatial slices or compactly supported spatial variations.

The induced metric, field strength \(F=\dd A\) and DBI matrix are
\begin{equation}
G_{ab}=g_{\mu\nu}\partial_aX^\mu\partial_bX^\nu,\qquad
F_{ab}=\partial_aA_b-\partial_bA_a,\qquad M=G+F.
\label{eq:dbi-data}
\end{equation}
Abou Zeid and Hull gave both a nonsymmetric auxiliary-tensor action linear in \(F\) and a separate auxiliary-metric action quadratic in \(F\) \cite{AbouZeidHull1997,AbouZeidHull1998}. Both recover the full DBI action by eliminating their auxiliary fields. We use the linear form with unit auxiliary normalization. Our tensor \(k\) transposes their convention so that contraction with \(M\) is an ordinary matrix trace. We take \(T>0\) on the real branch \(\det k<0\).
\begin{equation}
S_k=-\frac{T}{2}\int\dd^{p+1}\sigma\,\sqrt{-\det k}\,
       [\tr(k^{-1}M)-(p-1)].
\label{eq:parent}
\end{equation}
Vary \(k\) at fixed embedding and gauge potential. For \(Z_k=k^{-1}\delta k\), the determinant factor contributes \(\tr Z_k/2\) to its logarithmic variation, and the inverse tensor contributes \(-\tr(k^{-1}M Z_k)\) to the trace variation. All entries of \(Z_k\) are independent, so the stationary equation is
\begin{equation}
k^{-1}M=\frac{u}{2}\one,\qquad u=\tr(k^{-1}M)-(p-1).
\end{equation}
Taking the trace gives \(u=(p+1)u/2-(p-1)\), which fixes \(u=2\) for \(p>1\) and sets \(k_\star=M\). Substituting this solution into Eq.~\eqref{eq:parent} gives the DBI density \(-T\sqrt{-\det M}\). In varying the resulting action, the chain-rule term from \(k_\star\) vanishes by the auxiliary equation. The embedding and gauge equations therefore agree with those obtained from the auxiliary action. To prepare the time--space decomposition, use the inverse tensor density
\begin{equation}
Q=\sqrt{-\det k}\,k^{-1}
\end{equation}
gives the equivalent density
\begin{equation}
\mathcal L_Q=-\frac{T}{2}\bigl[\tr(QM)-(p-1)(-\det Q)^{1/(p-1)}\bigr].
\label{eq:parentQ}
\end{equation}
On the auxiliary solution, the time entry of \(Q\) is
\begin{equation}
(Q_\star)^{00}=-\frac{\det M_{\spat}}{\sqrt{-\det M}}<0.
\end{equation}
A positive-definite \(G_{\spat}\) ensures \(\det M_{\spat}>0\). Factoring out its square root reduces the determinant test to the identity plus a real antisymmetric matrix.

The time entry \(q\) and its Schur complement \(B_Q\) in \(Q\) define a positive lapse \(e\) and an independent spatial matrix \(s\). The symmetric and antisymmetric combinations of the normalized time--space blocks define the spatial shift \(N\) and an auxiliary vector \(W\), respectively.
\begin{equation}
Q=\begin{pmatrix}
q & qr^{\mathrm{T}}\\
ql & B_Q+qlr^{\mathrm{T}}
\end{pmatrix},\qquad
q=-\frac{1}{Te},\qquad B_Q=Te\,(\det s)s^{-1}.
\label{eq:schur}
\end{equation}
\begin{equation}
N=-\frac{l+r}{2},\qquad W=\frac{l-r}{2}.
\end{equation}
Here \(l,r\) are spatial columns. Row elimination gives \(\det Q=q\det B_Q\). Taking the determinant of the Schur block selects the positive root for \(\det s\) and makes the change of variables invertible for \(q<0\) and \(\det B_Q>0\).
\begin{equation}
\det s=\left(\frac{\det B_Q}{(Te)^p}\right)^{1/(p-1)},\qquad
(-\det Q)^{1/(p-1)}=Te\det s.
\end{equation}

The shifted embedding velocity and electric field, and the spatial auxiliary potential, are
\begin{equation}
U^\mu=\dot X^\mu-N^i\partial_iX^\mu,\qquad
\mathcal E_i=F_{0i}-N^jF_{ji},
\end{equation}
\begin{equation}
V(s,m)=\det s\,[\tr(s^{-1}m)-(p-1)],\qquad m_{\rm rel}=M_{\spat}.
\label{eq:potential}
\end{equation}
The argument \(m\) of \(V\) is a spatial DBI block, equal to \(m_{\rm rel}\) in the parent action. The trace separates into \(\tr(B_Qm_{\rm rel})\) and a term proportional to \(q\). Set \(l=-N+W\) and \(r=-N-W\). The symmetric part of the \(q\) term gives \(U^\mu U_\mu-W^iG_{ij}W^j\), and the antisymmetric part gives \(2W^i\mathcal E_i\). The Lagrangian density becomes
\begin{equation}
\mathcal L_Q=\frac{U^\mu U_\mu+2W^i\mathcal E_i-W^iG_{ij}W^j}{2e}
             -\frac{eT^2}{2}V(s,m_{\rm rel}).
\label{eq:blockl}
\end{equation}
The mixed equation and the momenta are
\begin{equation}
\mathcal E_i=G_{ij}W^j,\qquad
P_\mu=U_\mu/e,\qquad \pi^i=W^i/e .
\label{eq:velocity}
\end{equation}
The algebraic equation for \(W\) has a unique solution because \(G_{ij}\) is positive definite. It is convenient to set \(W=e\pi\), with \(\pi^i\) the momentum conjugate to \(A_i\). We call this electric displacement density the electric flux density. The linear \(W\) term becomes the gauge kinetic term, and the quadratic term becomes the electric contribution to the Hamiltonian. A Legendre transform in the embedding velocities and a spatial integration by parts in the \(A_0\) term give
\begin{equation}
S^{(1)}=\int\dd^{p+1}\sigma\,
  \bigl[P_\mu\dot X^\mu+\pi^i\dot A_i-\tfrac{e}{2}\mathcal C_{\rm rel}^{\rm aux}
  -N^i\mathcal C_i+A_0\mathcal G\bigr],
\label{eq:canonical}
\end{equation}
\begin{equation}
\mathcal C_{\rm rel}^{\rm aux}=P_\mu g^{\mu\nu}P_\nu+\pi^iG_{ij}\pi^j+T^2V(s,m_{\rm rel}),
\label{eq:finiteC}
\end{equation}
\begin{equation}
\mathcal C_i=P_\mu\partial_iX^\mu+\pi^jF_{ij},\qquad
\mathcal G=\partial_i\pi^i.
\label{eq:spatial-gauss}
\end{equation}
The lapse, shift and temporal potential impose the Hamiltonian (time), spatial diffeomorphism and Gauss constraints. The superscript \({\rm aux}\) indicates that \(s\) is independent. Setting \(s=m_{\rm rel}\) replaces the potential by \(\det m_{\rm rel}\) and recovers the canonical DBI constraints \cite{CorderoMolgadoRojas2007,Kluson2020}. The non-BPS constraints also involve the tachyon momentum, gradient and potential \cite{Kluson2017}. Appendix~\ref{app:direct} checks the constraints directly from the determinant. We now rescale the clock and gauge pairs with \(s\) still independent.

\section{Magnetic and electric Carroll contractions}
\label{sec:contractions}

\subsection{Common clock and gauge scalings}

Let \(c>0\) be the dimensionless contraction parameter and \(t\) the rescaled target time. We use \({\rm rel}\) for unscaled quantities and the subscript \(c\) for their expressions at finite \(c\) in rescaled variables. The clock, tension and lapse weights are common to both branches.
\begin{equation}
X_{\rm rel}^0=ct,\qquad P_t=cP_0^{\rm rel},\qquad
T_{\rm rel}=T_0/c,\qquad e_{\rm rel}=c^2e .
\label{eq:common-scaling}
\end{equation}
We hold the rescaled tension \(T_0=cT_{\rm rel}>0\) fixed in both gauge branches. With \(c=\omega^{-1}\), the tension and lapse weights agree with those used in non-BPS D-brane and canonical Carroll \(p\)-brane contractions \cite{Kluson2017,CardonaGomisPons2016}.

In the rescaled coordinates the time component of the flat target metric is \(-c^2\). As \(c\to0\), the metric degenerates along the clock direction. The target still has one clock direction and nine spatial directions, and the worldvolume dimension stays fixed. The clock rescaling is canonical because \(P_0^{\rm rel}\dd X_{\rm rel}^0=P_t\dd t\). The spatial embedding and its momenta, the shift and \(s\) are fixed in both branches. We scale the gauge potential \(A_a^{\rm rel}=c^{\alpha_{\rm g}}A_a\) and its spatial momentum \(\pi_{\rm rel}^i=c^{-\alpha_{\rm g}}\pi^i\). The opposite powers preserve the canonical kinetic term. We use \(\alpha_{\rm g}=0\) for the magnetic branch and \(\alpha_{\rm g}=1\) for the electric branch. Write \(h\) for the contribution of the nine spatial target directions to the induced metric. The induced spatial metric and the magnetic spatial block at finite parameter are
\begin{equation}
h_{ij}=\delta_{IJ}\partial_iX^I\partial_jX^J,\qquad
G_{c,ij}=h_{ij}-c^2\partial_it\,\partial_jt,\qquad m_{\M,c}=G_c+F_{\spat}.
\label{eq:spatialmetric}
\end{equation}
The rescaled lapse multiplies \(c^2\mathcal C_{\rm rel}^{\rm aux}/2\). In both branches the gauge weights cancel in the spatial generator and Gauss can be normalized using the rescaled flux.
\begin{equation}
\mathcal C_i=P_t\partial_i t+P_I\partial_iX^I+\pi^jF_{ij},\qquad
\mathcal G=\partial_i\pi^i.
\end{equation}
\subsection{Magnetic contraction}
\label{sec:magnetic}

In the magnetic branch the gauge potential and flux are fixed. The rescaled lapse factor cancels against the clock momentum and tension weights, leaving a suppression by \(c^2\) of both the spatial momentum and the electric flux squares. At fixed \(s\) the exact rescaled constraint is
\begin{equation}
\mathcal C_{\M,c}^{\rm aux}
  =-P_t^2+c^2P_IP_I+c^2\pi^iG_{c,ij}\pi^j+T_0^2V(s,m_{\M,c}).
\label{eq:cmc}
\end{equation}
Taking the limit in this Hamiltonian formulation gives
\begin{equation}
m_{\M}=h+F_{\spat},\qquad \mathcal C_{\M}^{\rm aux}=-P_t^2+T_0^2V(s,m_{\M}).
\label{eq:cmaux}
\end{equation}
To see how the change of variables absorbs the singular scaling, substitute the clock, tension and lapse weights into Eqs.~\eqref{eq:schur} and \eqref{eq:velocity}.
\begin{equation}
q=-\frac{1}{cT_0e},\qquad B_Q=cT_0e(\det s)s^{-1},\qquad
W=c^2w,\qquad w=e\pi .
\label{eq:block-scaling}
\end{equation}
With \(w\) independent, the linear term \(w^i\mathcal E_i/e\) in Eq.~\eqref{eq:blockl} fixes \(\pi=w/e\). The quadratic term \(-c^2w^iG_{c,ij}w^j/(2e)\) yields the suppressed flux square in Eq.~\eqref{eq:cmc}. The momentum of \(w\) vanishes and its bracket with the mixed relation has magnitude \(e^{-1}\). We call these relations the mixed pair. Their bracket is invertible at \(c=0\). Appendix~\ref{app:blocks} gives the limiting rank of the original tensor.

\subsection{Electric contraction}
\label{sec:electric}

The electric branch uses the same clock, tension and lapse weights and scales the gauge canonical pair as in the modified stringy contraction of Ref.~\cite{Kluson2017}.
\begin{equation}
A_a^{\rm rel}=c A_a,\qquad \pi_{\rm rel}^i=c^{-1}\pi^i .
\label{eq:escale}
\end{equation}
To check the mixed pair before its elimination, write \(W=cw\) and \(\mathcal E^{\rm rel}_i=c\mathcal E_i\) in Eq.~\eqref{eq:blockl}. Since \(e_{\rm rel}=c^2e\), the mixed sector becomes
\begin{equation}
\frac{w^i\mathcal E_i}{e}-\frac{w^iG_{c,ij}w^j}{2e}.
\label{eq:electric-mixed-block}
\end{equation}
The gauge momentum is \(\pi^i=w^i/e\). This relation and the vanishing momentum of \(w\) have an invertible mixed bracket of magnitude \(e^{-1}\) on the same positive-lapse domain as in the magnetic branch. Substituting \(w=e\pi\) gives the electric momentum square, while \(F_{ij}^{\rm rel}=cF_{ij}\) enters the auxiliary potential. The exact rescaled constraint is
\begin{equation}
\mathcal C_{\E,c}^{\rm aux}=-P_t^2+c^2P_IP_I+\pi^iG_{c,ij}\pi^j
                     +T_0^2V(s,m_{\E,c}),\qquad m_{\E,c}=G_c+cF_{\spat}.
\label{eq:cec}
\end{equation}
Taking the limit with \(s\) independent gives
\begin{equation}
m_{\E}=h,\qquad
\mathcal C_{\E}^{\rm aux}=-P_t^2+\pi^ih_{ij}\pi^j+T_0^2V(s,m_{\E}).
\label{eq:ceaux}
\end{equation}
The flux enters the time constraint and the spatial field strength still enters the spatial generator. The original auxiliary tensor has the same limiting rank as in the magnetic branch, as shown in Appendix~\ref{app:blocks}.

\subsection{Range of uniform gauge weights and Carroll boosts}

To compare the gauge weights, vary \(\alpha_{\rm g}\) at fixed clock, tension and lapse weights. The exponent includes any power absorbed into the DBI normalization. Hold the regular geometry, nonzero rescaled electric displacement and nonzero spatial field strength fixed. In target-time gauge the two gauge-dependent contributions to the reduced time constraint are
\begin{equation}
c^{2-2\alpha_{\rm g}}\pi^ih_{ij}\pi^j,\qquad
T_0^2\det(h+c^{\alpha_{\rm g}} F_{\spat}) .
\end{equation}
For the signed two-plane coefficients \(b_j\) of the antisymmetric field in an orthonormal spatial frame,
\begin{equation}
\det(h+c^{\alpha_{\rm g}} F_{\spat})=\det h\prod_j(1+c^{2\alpha_{\rm g}}b_j^2).
\label{eq:uniformdet}
\end{equation}
Finite clock energy requires \(0\leq\alpha_{\rm g}\leq1\). A smaller exponent makes the determinant diverge, and a larger exponent makes the electric term diverge. Both terms are nonnegative, so the divergence cannot cancel at fixed clock momentum. The two Carroll branches occur at the endpoints. The determinant retains its magnetic dependence at zero, and the electric displacement squared survives at one. In the interior only the spatial volume contributes at leading order. Both quadratic gauge corrections enter at order \(c\) when \(\alpha_{\rm g}=1/2\).

A target Carroll boost with constant parameter \(b_I^{\rm boost}\) shifts \(t\) and lifts canonically to the spatial momenta.
\begin{equation}
\delta t=b_I^{\rm boost}X^I,\qquad \delta P_I=-b_I^{\rm boost}P_t,\qquad
\delta X^I=\delta P_t=0 .
\end{equation}
The boost variations cancel in the canonical one-form and spatial constraint. The time constraints are invariant in both branches because the clock momentum, spatial fields and flux are invariant. The rank and symplectic reduction of the spatial auxiliary pair are therefore the next steps for both theories.

\section{Hamiltonian phase space and Dirac reduction}
\label{sec:reduction}

On an equal-time slice \(\Sigma\), the auxiliary constraint surface determines the pullback of the symplectic form to the matter variables. Restricting to the first-class constraint surface and identifying points along its gauge flows then gives the local physical phase space.

\subsection{Rank of the spatial auxiliary pair}

We enforce preservation of the constraints under Hamiltonian evolution, following Dirac's stabilization procedure \cite{Dirac1950,Dirac1964}. The momentum \(\rho^{ij}\) conjugate to \(s_{ij}\) vanishes because the action contains no time derivative of that tensor. Preserving this primary constraint gives the secondary equation for \(s_{ij}\).
\begin{equation}
\rho^{ij}\simeq0,\qquad
\psi^{ij}=\frac{\partial V}{\partial s_{ij}}\simeq0.
\end{equation}
The primary evolves as \(\dot\rho^{ij}=-eT_0^2\psi^{ij}/2\), so the positive lapse and tension permit division by its prefactor. The rank calculation uses a general spatial block \(m\). At the Carroll endpoint it is \(m_{\M}\) in the magnetic branch and \(m_{\E}=h\) in the electric branch of Sec.~\ref{sec:electric}. At fixed matter fields, let \(Z_s=s^{-1}\delta s\). Then
\begin{equation}
\delta V=\det s\,\tr[(u\one-s^{-1}m)Z_s],\qquad
u=\tr(s^{-1}m)-(p-1).
\end{equation}
Nondegeneracy of the trace pairing on full real matrices gives \(s^{-1}m=u\one\). Taking the trace fixes \(u=1\), hence \(s_\star=m\) and \(V(s_\star,m)=\det m\).

The Hessian follows by linearizing \(u\one-s^{-1}m\) at \(s=m\). At that point \(\delta u=-\tr Z_s\) and \(\delta(s^{-1}m)=-Z_s\). The stationary equation removes terms from differentiating the prefactors or choosing a field-dependent parametrization. For two independent variations with \(Z_s=m^{-1}\delta s\), the bilinear form is
\begin{equation}
H_s(\delta s,\delta' s)=\det m\,[\tr(Z_sY_s)-\tr Z_s\,\tr Y_s],\qquad
Y_s=m^{-1}\delta' s .
\label{eq:hessian}
\end{equation}
The trace map \(Z_s\mapsto Z_s-(\tr Z_s)\one\) has eigenvalue one on traceless matrices and \(1-p\) on the trace direction. Both are nonzero for \(p>1\). The trace pairing on unrestricted matrices and the nonsingular \(m\) therefore make \(H_s\) invertible. The auxiliary bracket matrix is
\begin{equation}
\Delta_{\rm aux}=\begin{pmatrix}0&-H_s\\ H_s&0\end{pmatrix}.
\label{eq:diracmatrix}
\end{equation}
Each entry multiplies the spatial delta function. The secondary constraints Poisson commute because they depend only on configurations. The invertible off-diagonal block gives rank \(2p^2\) and fixes the primary multipliers through stabilization.

We call \((s,\rho)=(m,0)\) the spatial auxiliary constraint surface. On this surface, the auxiliary term \(\int_\Sigma\rho^{ij}\delta s_{ij}\) vanishes. The pulled-back potential is
\begin{equation}
\vartheta_{\rm mat}=\int_\Sigma\dd^p\sigma\,
                 (P_t\delta t+P_I\delta X^I+\pi^i\delta A_i).
\label{eq:theta}
\end{equation}
Its exterior derivative gives the canonical matter symplectic form. Substituting the stationary potential into the magnetic and electric constraints gives the following time densities. The positive root of the electric constraint defines \(\varepsilon_{\E}\).
\begin{equation}
\mathcal C_{\M}=-P_t^2+T_0^2\det(h+F_{\spat}).
\label{eq:cm}
\end{equation}
\begin{equation}
\mathcal C_{\E}=-P_t^2+T_0^2\det h+\pi^ih_{ij}\pi^j,\qquad
\varepsilon_{\E}=\sqrt{T_0^2\det h+\pi^ih_{ij}\pi^j}.
\label{eq:ce}
\end{equation}

\subsection{Residual constraint algebra and its contraction}

For either branch let \(a\in\{\M,\E\}\) label the time density. Use the canonical matter bracket and smear the constraints by \(\alpha,\xi,f\).
\begin{equation}
H[f]=\tfrac12\int_\Sigma f\mathcal C_a,\qquad
D[\xi]=\int_\Sigma \xi^i(\mathcal C_i-A_i\mathcal G),\qquad
\mathcal G[\alpha]=-\int_\Sigma\alpha\mathcal G .
\label{eq:generators}
\end{equation}
The Gauss term in \(D[\xi]\) supplies the exact differential in \(\mathcal L_\xi A=\iota_\xi F+\dd(\iota_\xi A)\), so \(D[\xi]\) generates ordinary Lie derivatives. The gauge-covariant representative has additional Gauss terms in its brackets, as in the related non-BPS algebra \cite{ZengPhase2026}. On the common constraint surface either representative, together with Gauss, generates the same local gauge distribution. The test function \(f\) used to smear the time constraint has density weight minus one, since the constraint has density weight two. Integration by parts gives
\begin{equation}
\{D[\xi],D[\eta]\}=D[[\xi,\eta]],\qquad
\{D[\xi],H[f]\}=H[\mathcal L_\xi f],\qquad
\{D[\xi],\mathcal G[\alpha]\}=\mathcal G[\mathcal L_\xi\alpha].
\end{equation}
The limiting time generators commute strongly, meaning that their bracket vanishes throughout the reduced matter phase space. In the magnetic density the variables \(P_t,X^I,A_i\) and their spatial derivatives commute mutually. In the electric density the corresponding commuting set uses \(\pi^i\) in place of \(A_i\). Gauss changes \(A_i\) by a gradient and leaves \(F_{ij}\) and \(\pi^i\) invariant. The remaining brackets among \(H\) and \(\mathcal G[\alpha]\) therefore vanish strongly in both branches. The full multiplier chains are given below.

\paragraph{Finite-parameter time bracket.}
The relativistic DBI constraint algebra has a field-dependent coefficient in the bracket of two Hamiltonian constraints \cite{CorderoMolgadoRojas2007}. We derive this coefficient in our conventions to track its scaling. In unscaled variables, \(H_{\rm rel}[f]=\frac12\int_\Sigma f\mathcal C_{\rm rel}\), with \(\mathcal C_{\rm rel}=\mathcal C_{\rm rel}^{\rm aux}|_{s=m_{\rm rel}}\). Set \(m_{\rm rel}=G_{\spat}+F_{\spat}\), \(d_{\rm rel}=\det m_{\rm rel}\) and \(R^{ij}=(m_{\rm rel}^{-1})^{ji}\). The gradient derivatives of the Hamiltonian density are
\begin{equation}
J_\mu^i=\mathcal S^{ij}\partial_jX_\mu,\qquad
K^{ij}=T^2d_{\rm rel} R^{[ij]},\qquad
\mathcal S^{ij}=\pi^i\pi^j+T^2d_{\rm rel} R^{(ij)} .
\label{eq:finite-gradient-currents}
\end{equation}
These follow from \(\delta d_{\rm rel}=d_{\rm rel} R^{ij}\delta(m_{\rm rel})_{ij}\). The symmetric part multiplies the induced-metric variation. The antisymmetric part multiplies the field-strength variation. Integrating by parts, the derivatives of \(H_{\rm rel}[f]\) with respect to \(X^\mu\) and \(A_j\) are \(-\partial_i(fJ_\mu^i)\) and \(-\partial_i(fK^{ij})\). Its momentum derivatives are \(fP^\mu\) and \(fG_{jk}\pi^k\). In their canonical bracket, antisymmetrizing in \(f,g\) cancels all derivatives of the currents, leaving
\begin{equation}
\{H_{\rm rel}[f],H_{\rm rel}[g]\}
=\int_\Sigma b_i\bigl(J_\mu^iP^\mu+K^{ij}G_{jk}\pi^k\bigr),
\qquad b_i=f\partial_i g-g\partial_i f .
\end{equation}
Let \(R_{\rm a}\) and \(R_{\rm s}\) be the antisymmetric and symmetric parts of \(R\). Subtracting \(R(G_{\spat}-F_{\spat})=\one\) and \(R^{\mathrm{T}}(G_{\spat}+F_{\spat})=\one\) gives \(R_{\rm a}G_{\spat}=R_{\rm s}F_{\spat}\). Also \(\pi^iF_{ij}\pi^j=0\). The two currents thus combine into the covariant spatial constraint.
\begin{equation}
\{H_{\rm rel}[f],H_{\rm rel}[g]\}=C[v],\qquad
v^j=\mathcal S^{ij}b_i,\qquad C[v]=\int_\Sigma v^i\mathcal C_i .
\label{eq:finite-time-algebra}
\end{equation}
With Eq.~\eqref{eq:generators}, \(C[v]=D[v]+\mathcal G[-v^iA_i]\). The time smearings are field independent. Gauge invariance and Lie differentiation give the remaining brackets as in the limiting calculation. Appendix~\ref{app:bosonic-time-details} gives the intermediate derivatives and the leading coefficients of the contraction.

In the rescaled canonical variables \(H_c[f]=c^2H_{\rm rel}[f]\). The spatial constraint keeps the same form. Use the two gauge weights of Sec.~\ref{sec:contractions}. The structure function in Eq.~\eqref{eq:finite-time-algebra} is multiplied by \(c^4\). Using \(T_{\rm rel}=T_0/c\) gives
\begin{equation}
\mathcal S_c^{ij}
=c^{4-2\alpha_{\rm g}}\pi^i\pi^j
+c^2T_0^2\det(m_{\alpha_{\rm g},c})(m_{\alpha_{\rm g},c}^{-1})^{(ij)},\qquad
m_{\alpha_{\rm g},c}=G_c+c^{\alpha_{\rm g}} F_{\spat}.
\label{eq:contracted-time-structure}
\end{equation}
For fixed smooth smearings and bounded rescaled data, the finite symmetric cofactor gives an \(O(c^2)\) bracket in both branches, consistent with strong commutativity at \(c=0\). The flux contribution starts at \(O(c^4)\) in the magnetic branch and \(O(c^2)\) in the electric branch. The limiting time and Gauss generators leave \(X^I\) invariant. Because the embedding gradients have full spatial rank, a vanishing combined flow must have zero spatial smearing. Its time smearing then vanishes since \(P_t\ne0\) on the positive-tension time surface. Gauss acts by a gradient, so its smearing is fixed locally up to a constant. The \(p+2\) independent local first-class constraints reduce the \(10+p\) matter pairs to eight physical pairs. This includes pairs whose configurations freeze when worldvolume time equals target time. Constant Gauss smearing gives a global identity on a closed slice.

\subsection{Complete stabilization and local degrees of freedom}
\label{sec:dirac-completeness}

Before eliminating the mixed vector, treat \(w^i\) as an independent coordinate. Its momentum \(p_{w,i}\) and the gauge momentum relation
\(\chi^i=\pi^i-w^i/e\) are primary constraints in both rescaled branches. At finite parameter the same equations use the unscaled \(W,e,\pi\). The embedding velocities, or the supervielbein velocities at fixed fermionic velocity, have an invertible Legendre map for nonzero \(e\). The gauge velocities enter linearly at fixed \(w\), so their only spatial primary relations are \(\chi^i=0\). The primary brackets are
\begin{equation}
\{p_{w,i}(x),\chi^j(y)\}=e^{-1}\delta_i^{\,j}\delta_\Sigma(x,y),
\qquad
\{p_{w,i},p_{w,j}\}=\{\chi^i,\chi^j\}=0 .
\label{eq:mixed-primary-bracket}
\end{equation}
Preserving these \(2p\) relations determines their \(2p\) multipliers. In particular, \(\chi^i\) depends on the lapse, so the bare lapse momentum need not commute with it. The combination \(p_e+(w^i/e)p_{w,i}\) has zero bracket with the mixed pair on its constraint surface. On that surface, defined by \(p_w=0,\ w=e\pi\), the term \(p_{w,i}\delta w^i\) pulls back to zero. The lapse pair is then canonical on the induced symplectic form.

After this step the remaining even primary constraints are \(\rho^{ij}=0\) and the momenta \(p_e,p_{N,i},\pi^0\) of \(e,N^i,A_0\). For either regular branch the Hamiltonian on the mixed constraint surface has the form
\begin{equation}
H_{\rm tot}=\int_\Sigma\left(
 \tfrac e2\mathcal C^{\rm aux}+N^i\mathcal C_i-A_0\mathcal G
 +u_{ij}\rho^{ij}+u_ep_e+u_N^ip_{N,i}+u_0\pi^0\right).
\label{eq:total-even-hamiltonian}
\end{equation}
Here the unscaled parent has \(T\) in its auxiliary potential and the rescaled theories have \(T_0\). Preserving \(p_e,p_{N,i},\pi^0\) gives \(-\mathcal C^{\rm aux}/2,-\mathcal C_i,\mathcal G\), respectively, as their time derivatives. Preserving \(\rho\) gives the spatial stationarity constraints \(\psi\). In collective auxiliary indices \(A,B\), the next condition is \(\{\psi^A,H_{\rm tot}|_{u_{ij}=0}\}+H_s^{AB}u_B\simeq0\). Since \(H_s\) is invertible, this determines \(u_B\) and produces no further condition. All these steps apply before and after contraction.

On the spatial auxiliary constraint surface the remaining \(p+2\) secondary constraints are the time, spatial and Gauss constraints of the reduced matter theory. The preceding subsection establishes their algebra for both branches. The \(p+2\) multiplier momenta commute with the reduced constraints, which are independent of \(e,N^i,A_0\). Thus preservation of the secondary constraints imposes no tertiary constraint and leaves the first-class multipliers arbitrary. To represent the first-class constraints on a phase space with a second-class set \(\chi_A\) still present, use
\begin{equation}
F^\star=F-\{F,\chi_A\}(\Delta^{-1})^{AB}\chi_B,
\qquad \Delta_{AB}=\{\chi_A,\chi_B\},
\label{eq:first-class-lift}
\end{equation}
where repeated field-theory indices include integration. The bracket of \(F^\star\) with \(\chi_A\) vanishes on the second-class constraint surface. The Poisson bracket of two such extensions equals their Dirac bracket on the second-class surface \cite{Dirac1964,HenneauxTeitelboim1992}. This applies successively to the mixed and spatial sets and explains the status of the multiplier momenta before elimination.

Independence of the finite generators follows from their action on the embedding and gauge field. The finite time constraint makes the embedding momentum timelike since the flux square is nonnegative and the spatial determinant term is positive. The spatial embedding gradients span a spacelike subspace, so a vanishing combined embedding flow forces both the time and spatial smearings to vanish. The remaining Gauss flow vanishes locally only for constant smearing. The limiting independence argument was given above. Including the multiplier pairs, the even phase space has \(10+p+(p+p^2)+(p+2)\) canonical pairs before imposing constraints. There are \(2(p+p^2)\) second-class constraints and \(2(p+2)\) first-class constraints. The local physical count is eight canonical pairs. The count treats the constant Gauss smearing on a closed slice as a global identity and uses only independent local generators. At finite parameter the Schur change of variables is invertible, so this list also accounts for the primary relations of the original auxiliary action on the same real branch.

\subsection{Commutation of auxiliary elimination and contraction}
\label{sec:order-of-reduction}

For each gauge exponent in Eq.~\eqref{eq:contracted-time-structure}, write \(m_{\alpha_{\rm g}}=\lim_{c\to0}m_{\alpha_{\rm g},c}\). The mixed pair remains invertible in both branches by Secs.~\ref{sec:magnetic} and \ref{sec:electric}. Eliminating the spatial pair at finite \(c\) sets \((s,\rho)=(m_{\alpha_{\rm g},c},0)\) and gives the potential \(\det m_{\alpha_{\rm g},c}\). Contracting first gives \(V(s,m_{\alpha_{\rm g}})\), with stationary surface \((s,\rho)=(m_{\alpha_{\rm g}},0)\). Thus both orders yield the same limiting potential \(\det m_{\alpha_{\rm g}}\).

Both constraint surfaces lie in a common regular region. In a spatial reference norm, bound the eigenvalues of \(h\) away from zero and keep \(\partial_i t\) bounded. Then \(G_c\) stays uniformly positive for small \(c\). The antisymmetric part of \(m_{\alpha_{\rm g},c}\) drops out of the real quadratic form. Cauchy--Schwarz therefore bounds the smallest singular value of \(m_{\alpha_{\rm g},c}\) below by the smallest eigenvalue of \(G_c\). With bounded rescaled data and lapse bounded away from zero, the auxiliary constraint surface is smooth and the inverse Hessian in Eq.~\eqref{eq:hessian} stays bounded.

The auxiliary momenta vanish in either order, giving the same pullback \eqref{eq:theta}. The limiting time, spatial and Gauss densities agree as well. The identity on the matter variables therefore maps each residual Hamiltonian flow to itself and induces a local symplectomorphism between the two reduced phase spaces.

The common reduction gives two different Hamiltonians in target-time gauge. We use the same gauge and initial-data conditions to compare their dynamics in Sec.~\ref{sec:examples}.

\section{Bosonic dynamics in target-time gauge}
\label{sec:examples}

\subsection{Energy and evolution in target-time gauge}

In both branches target-time gauge sets the target time \(t\) equal to the worldvolume time \(\tau\). The bracket of \(t\) with either half time density is \(-P_t\) times the spatial delta function. Preserving the gauge gives \(-eP_t=1\), so positive lapse chooses the negative root. The electric energy is given in Eq.~\eqref{eq:ce}. For the magnetic branch the result is
\begin{equation}
t=\tau,\qquad P_t=-\varepsilon_{\M},\qquad
e=\varepsilon_{\M}^{-1},\qquad \varepsilon_{\M}=T_0\sqrt{\det m_{\M}}.
\label{eq:hphysm}
\end{equation}
Substituting the clock momentum into the action gives \(H^{\rm phys}_a=\int_\Sigma\varepsilon_a\), with the spatial and Gauss constraints still imposed. In a local gauge with zero shift and temporal potential, the spatial embeddings freeze in both branches, since neither Hamiltonian depends on \(P_I\). In the magnetic branch the gauge potential also freezes because the Hamiltonian is independent of \(\pi^i\). To obtain the momentum equations, vary the determinant using \(\delta\sqrt{\det m_{\M}}=\sqrt{\det m_{\M}}\,(m_{\M}^{-1})^{ji}\delta (m_{\M})_{ij}/2\). The induced metric variation contracts with its symmetric part and the field-strength variation with its antisymmetric part. With \(R^{ij}=(m_{\M}^{-1})^{ji}\), integration by parts gives
\begin{equation}
\dot P_I=T_0\partial_i\bigl(\sqrt{\det m_{\M}}\,R^{(ij)}\partial_jX_I\bigr),\qquad
\dot\pi^{j}=T_0\partial_i\bigl(\sqrt{\det m_{\M}}\,R^{[ij]}\bigr).
\label{eq:force}
\end{equation}
The divergence of the gauge force vanishes by antisymmetry, preserving Gauss. These momentum equations show how a frozen configuration can still exert finite stresses.

In the electric branch the Hamiltonian is independent of \(A_i\), so \(\dot\pi^i=0\), while its flux derivative gives \(\dot A_i=h_{ij}\pi^j/\varepsilon_{\E}\). At fixed flux its metric variation is
\begin{equation}
\delta\varepsilon_{\E}=\frac{\mathcal S_{\E}^{ij}}{2\varepsilon_{\E}}\delta h_{ij},
\qquad
\mathcal S_{\E}^{ij}=T_0^2\det h\,(h^{-1})^{ij}+\pi^i\pi^j.
\end{equation}
Inserting \(\delta h_{ij}=2\partial_{(i}X_I\partial_{j)}\delta X^I\) and integrating by parts gives the corresponding embedding force.
\begin{equation}
\dot P_I=\partial_i\left(\frac{\mathcal S_{\E}^{ij}}{\varepsilon_{\E}}\partial_jX_I\right).
\label{eq:electric-force}
\end{equation}
The embedding momentum can therefore evolve in both branches at fixed spatial embedding. The same separation between a static embedding and evolving momentum occurs for Carroll domain walls \cite{ClarkterVeldhuis2016}. Here the spatial gauge field and the electric displacement supply the additional stresses in their respective branches.

\subsection{Finite-parameter continuation and homogeneous D2 solutions}

The rescaled Hamiltonian formulation gives an explicit continuation of equal-time data to finite \(c\) in the magnetic branch. Keep the matter data that satisfy the spatial and Gauss constraints. In target-time gauge \(\partial_i t=0\), so \(m_{\M,c}=m_{\M}\). Solving the finite-parameter time constraint gives
\begin{equation}
\varepsilon_c=\sqrt{T_0^2\det m_{\M}+c^2(P_IP_I+\pi^ih_{ij}\pi^j)},\qquad
P_t(c)=-\varepsilon_c,\qquad e(c)=\varepsilon_c^{-1}.
\label{eq:constraintcontinuation}
\end{equation}
These data satisfy the finite-parameter constraints and preserve target-time gauge. For bounded smooth data with a positive lower bound on the DBI volume, \(\varepsilon_c-\varepsilon_{\M}=O(c^2)\) at equal time. In the electric branch the same construction at fixed matter data uses
\begin{equation}
\varepsilon_{\E,c}=
\sqrt{c^2P_IP_I+\pi^ih_{ij}\pi^j+T_0^2\det(h+cF_{\spat})},
\qquad
P_t(c)=-\varepsilon_{\E,c},\qquad e(c)=\varepsilon_{\E,c}^{-1}.
\label{eq:electric-continuation}
\end{equation}
Transposition gives \(\det(h+cF_{\spat})=\det(h-cF_{\spat})\). The determinant is polynomial in \(c\), so the same positive lower bound on the energy gives \(\varepsilon_{\E,c}-\varepsilon_{\E}=O(c^2)\). These are equal-time continuations. The homogeneous examples below give exact trajectories for both branches.

\paragraph{Magnetic branch.}
Take a unit planar D2 in static spatial gauge, with \(X^1=\sigma^1\) and \(X^2=\sigma^2\). Keep one transverse coordinate \(Z\), with constant momentum \(P_Z\). Give the brane a constant electric flux \(\pi\) and magnetic field \(F_{12}=B\). The spatial constraint fixes \(P_\parallel=-F\pi\), and constant flux satisfies Gauss. Since \(F^{\mathrm{T}}F=B^2\one\), the tangential momentum square and electric flux square combine into \(\mathcal D\,\pi^{\mathrm{T}}\pi\), where \(\mathcal D=1+B^2\). The time constraint and gauge-preservation equations then give
\begin{equation}
\varepsilon_c=\sqrt{T_0^2\mathcal D+c^2(P_Z^2+\mathcal D\,\pi^{\mathrm{T}}\pi)},\qquad
P_t=-\varepsilon_c,\qquad e=\varepsilon_c^{-1},
\end{equation}
\begin{equation}
N=-c^2eP_\parallel,\qquad
\dot Z=c^2eP_Z,\qquad F_{0i}=c^2e\mathcal D\,\pi_i .
\label{eq:d2vel}
\end{equation}
The shift in Eq.~\eqref{eq:d2vel} preserves the planar gauge and combines with the direct electric term to give the factor \(\mathcal D\) in \(F_{0i}\). The spatial currents are constant, so their divergences vanish and the momenta remain constant. Substitution in the full DBI matrix gives
\begin{equation}
\sqrt{-\det M}=cT_0e\mathcal D .
\end{equation}
Differentiating the DBI density reproduces the transverse, clock and gauge momenta on the real finite-parameter branch. On bounded time intervals the configuration velocities vanish quadratically in \(c\), while the canonical electric flux remains finite.

\paragraph{Electric branch.}
Keep the same planar data \(B,\pi,P_Z\) and use the electric weights. The spatial constraint still fixes \(P_\parallel=-F\pi\). With \(\mathcal D_c=1+c^2B^2\), Eq.~\eqref{eq:electric-continuation} and gauge preservation give
\begin{equation}
\begin{gathered}
\varepsilon_{\E,c}^{\,2}=\mathcal D_c(T_0^2+\pi^{\mathrm T}\pi)+c^2P_Z^2,
\qquad P_t=-\varepsilon_{\E,c},\qquad e=\varepsilon_{\E,c}^{-1},\\
N=-c^2eP_\parallel,\qquad \dot Z=c^2eP_Z,\qquad
F_{0i}=e\pi_i+N^jF_{ji}=e\mathcal D_c\pi_i .
\end{gathered}
\label{eq:electric-planar}
\end{equation}
The currents are spatially constant, so all momenta remain constant. The Schur complement of \(\one+cF_{\spat}\) in the DBI matrix gives
\begin{equation}
-\det M/c^2=\mathcal D_c(\dot t^{\,2}-\dot Z^{\,2}/c^2)-F_{0i}F_{0i}
\end{equation}
in static spatial gauge. Before imposing that gauge, a tangent velocity \(u\) contributes \(2u^{\mathrm T}F F_0\) at first order to this expression, where \(F_0\) is the column with entries \(F_{0i}\). Differentiation of the DBI density and substitution of Eq.~\eqref{eq:electric-planar} yield \(\sqrt{-\det M}=cT_0e\mathcal D_c\) and the same clock, transverse, gauge and tangent momenta. Thus the rescaled gauge velocity remains finite for nonzero flux. When \(B=P_Z=0\), the energy is independent of \(c\) and the DBI volume is \(cT_0/\varepsilon_{\E}\).

\subsection{Inhomogeneous D2 solutions and constraint propagation}

\paragraph{Magnetic branch.}
On a spatial patch use the same planar embedding and let \(B=B(x)\) be smooth. Write \(\mathcal D(x)=1+B(x)^2\), take a fixed potential with \(\partial_x A_2=B\), and set \(\pi^1=P_2=P_Z=0\). Equations~\eqref{eq:hphysm} and \eqref{eq:force} then give
\begin{equation}
P_t=-T_0\sqrt{\mathcal D},\qquad
\pi^2(\tau,x)=\pi^2(0,x)+\tau T_0\,\frac{B'(x)}{\mathcal D^{3/2}},
\qquad P_1=-B\pi^2 .
\label{eq:inhomogeneous}
\end{equation}
The two forces are \(\dot\pi^2=T_0\partial_x(B/\sqrt{\mathcal D})\) and \(\dot P_1=T_0\partial_x(\mathcal D^{-1/2})\). They satisfy \(\dot P_1+B\dot\pi^2=0\), preserving the spatial constraint. Gauss holds because \(\pi^2\) is independent of the second coordinate. The flux evolves at fixed magnetic field and energy density.

\paragraph{Electric branch.}
In the electric branch keep the same planar embedding and take \(\pi^x=0\), \(\pi^y=j(x)\) with smooth \(j\). The flux and \(\varepsilon_{\E}(x)=\sqrt{T_0^2+j(x)^2}\) are static. From zero initial gauge potential the gauge velocity integrates to
\begin{equation}
A_x=0,\qquad A_y=\tau\, j/\varepsilon_{\E},\qquad
B(\tau,x)=\tau T_0^2j'(x)/\varepsilon_{\E}^{\,3}.
\label{eq:electric-inhomogeneous}
\end{equation}
Choose \(P_y=P_Z=0\) and \(P_x=-jB\). Gauss holds because the flux is independent of \(y\). Equation~\eqref{eq:electric-force} gives \(\dot P_x=\partial_x(T_0^2/\varepsilon_{\E})=-j\dot B\), so the spatial constraint is preserved. Thus the inhomogeneous magnetic field drives the flux in the magnetic solution, and the inhomogeneous electric flux generates magnetic curvature in the electric solution.

\subsection{Nonlinear magnetic energy and electric line tension}

\paragraph{Two magnetic planes on a D4.}
Take a planar D4 with unit induced spatial metric and constant fields \(F_{12}=B_1\) and \(F_{34}=B_2\). Choose zero spatial momenta, zero electric flux and zero shift. The auxiliary equation fixes \(s=\one+F_{\spat}\), and the reduced constraint gives
\begin{equation}
\varepsilon_{\M}^{\rm D4}=T_0\sqrt{(1+B_1^2)(1+B_2^2)} .
\label{eq:d4}
\end{equation}
Each magnetic plane contributes its own \(2\times2\) block to the determinant. In the direct DBI action the time block is \(-c^2\dot t^2\), so on the positive-clock branch the density is \(-T_0\dot t\sqrt{(1+B_1^2)(1+B_2^2)}\). Its derivative with respect to \(\dot t\) reproduces the clock momentum from the constraint. The spatial constraints vanish and the stress is constant.

The coefficient of \(B_1^2B_2^2\) in the time constraint at fixed \(P_t\) is \(T_0^2\). For the physical energy, each square-root factor contributes a quadratic coefficient \(1/2\). Multiplying the two factors therefore gives the mixed Taylor coefficient \(T_0/4\) in Eq.~\eqref{eq:d4}.

\paragraph{Flux-aligned D2 in flat space.}
\label{sec:flux-lines}

A relativistic D-brane with constant electric field can carry fundamental-string charge \cite{DiVecchiaFrauLerdaLiccardo2000}. In the tachyon vacuum the flux describes a string fluid, and a flux-aligned ansatz gives a string sector in the corresponding non-BPS Carroll theory \cite{GibbonsHoriYi2001,Kluson2017,Kluson2022}. We use that alignment to compute the force at fixed positive brane tension and without a tachyon. Let \(x,y\) be spatial worldvolume coordinates, with a prime denoting \(\partial_x\). Take a D2 along the curve \(\boldsymbol X(x)=(x,Z(x))\) and extend it along \(X^2=y\), with the other fields independent of \(y\). We refer to this straight direction as the spectator direction. In target-time gauge with zero shift and temporal potential, set \(\pi^x=J\) constant, \(\pi^y=0\) and initially \(F_{xy}=0\). Then \(\det h=h_{xx}=1+Z'^2\). Equation~\eqref{eq:ce} gives the energy per coordinate area and the effective line tension, the coefficient of the curve length per unit length in the straight direction.
\begin{equation}
\varepsilon_{\rm E}=\sqrt{h_{xx}}\,\mathcal T,\qquad
\mathcal T=\sqrt{T_0^2+J^2}.
\label{eq:flux-line-energy}
\end{equation}
The embedding and flux are time independent. The gauge equation gives \(\dot A_x=J\sqrt{h_{xx}}/\mathcal T\) and \(\dot A_y=0\), so \(F_{xy}\) stays zero. The embedding momentum in the curve plane evolves according to
\begin{equation}
\dot{\boldsymbol P}=\mathcal T\,
\partial_x\left(\frac{\partial_x\boldsymbol X}{\sqrt{h_{xx}}}\right).
\label{eq:flux-line-force}
\end{equation}
The stress current has tangent coefficient \((T_0^2+J^2)/\varepsilon_{\rm E}=\mathcal T/\sqrt{h_{xx}}\). Its divergence gives the force in Eq.~\eqref{eq:flux-line-force}. The force is normal, so an initially vanishing tangential momentum stays zero. The spectator momentum is constant and Gauss holds since \(J\) is constant. For a periodic spectator of length \(L_y\), the integrated energy is \(L_y\mathcal T\int\dd x\,\sqrt{h_{xx}}\). The corresponding magnetic force at zero spatial field strength has coefficient \(T_0\).

\subsection{D2 curves in a static warped Carroll background}
\label{sec:warped-flux}

Flat space isolates the stress due to the embedding and gauge data. To examine background gradients, place both branches in a static warped product with an orthogonal spatial Killing direction \(Y\). This metric is a specialization of the small-\(c\) expansion of Carroll geometry used in the D-brane analysis of Ref.~\cite{HansenObersOlingSogaard2022,Kluson2022}.
\begin{equation}
ds_c^2=-c^2a(X)^2dt^2+\gamma_{mn}(X)dX^mdX^n+r_Y(X)^2dY^2 .
\label{eq:warped-metric}
\end{equation}
Let \(a,r_Y>0\) and take the base metric \(\gamma\) to be Riemannian. The background two-form vanishes and the dilaton is constant. Write \(h\) for the pullback of the spatial metric in Eq.~\eqref{eq:warped-metric}. At finite \(c\) the spatial auxiliary solution in each branch is \(h_{ij}-c^2a^2\partial_it\,\partial_jt+c^{\alpha_{\rm g}}F_{ij}\). Its symmetric part stays positive on the same regular domain. The inverse clock metric \(g^{tt}=-1/(c^2a^2)\) gives \(-P_t^2/a^2\) in both limiting densities. Preserving \(t=\tau\) thus gives \(e=a^2/\varepsilon\), where \(\varepsilon=-P_t>0\) is the energy of the branch in question. The static background depends only on spatial embeddings, so the time bracket in either branch still vanishes strongly. Consider a curve \(X^m=X^m(x)\) extended along \(Y=y\), with all other data independent of \(y\). Set the shift and temporal potential to zero. The induced matrix is \(\diag(h_{xx},r_Y^2)\), with \(h_{xx}=\gamma_{mn}X'^mX'^n\).

\paragraph{Magnetic branch.}
Take \(F_{xy}=B(x)\) and \(\pi^x=0\). Both the spatial embedding and the magnetic field are static. Writing \(d=r_Y^2h_{xx}+B^2\), the time constraint has potential \(T_0^2d\) and gives \(\varepsilon_{\M}=aT_0\sqrt d\). Vary before substituting the embedding. The symmetric part of the inverse spatial DBI matrix is diagonal on this configuration, so the \(Y\)-variation gives a current only along \(y\). The isometry and \(y\)-independence then give \(\dot P_Y=0\). Base and gauge variations use \(\delta d=h_{xx}\delta(r_Y^2)+r_Y^2\delta h_{xx}+2B\delta B\). Integration by parts gives
\begin{equation}
\begin{gathered}
U_m=\frac{aT_0r_Y^2\gamma_{mn}X'^n}{\sqrt d},
\qquad I_B=\frac{aT_0 B}{\sqrt d},\\
\dot P_m=\partial_x U_m-\left.\partial_m\varepsilon_{\M}\right|_{X',B},
\qquad \dot\pi^y=\partial_x I_B,\qquad\dot\pi^x=0 .
\end{gathered}
\label{eq:warped-magnetic-force}
\end{equation}
Here \(U_m\) and \(I_B\) are the derivatives of the energy density with respect to \(X'^m\) and \(B\). Since \(X'^mU_m+B I_B=\varepsilon_{\M}\), differentiating this identity and applying the chain rule gives \(X'^m\dot P_m+B\dot\pi^y=0\). Thus \(P_mX'^m+B\pi^y=0\) is preserved. The other spatial constraint is \(P_Y-B\pi^x=0\) and Gauss holds by \(y\)-independence. The static background preserves the clock momentum.

\paragraph{Electric branch.}
Set \(\pi^x=J\) constant and \(\pi^y=0\), with zero initial \(F_{xy}\). The reduced time constraint and its energy are
\begin{equation}
\mathcal C_{\E}=-\frac{P_t^2}{a^2}+T_0^2\det h+\pi^ih_{ij}\pi^j,\qquad
\varepsilon=a\sqrt{T_0^2\det h+\pi^ih_{ij}\pi^j}.
\label{eq:warped-clock}
\end{equation}
Per unit coordinate period of the spectator, the energy is a weighted curve length with
\begin{equation}
\varepsilon=\sqrt{h_{xx}}\,\mathcal T(X),\qquad
\mathcal T(X)=a(X)\sqrt{T_0^2r_Y(X)^2+J^2}.
\label{eq:warped-line}
\end{equation}
The embedding and flux are time independent. The gauge velocity has only an \(x\)-component, independent of \(y\), so \(F_{xy}=0\) persists. Varying the full Hamiltonian in target-time gauge with respect to \(Y\) before imposing the ansatz gives a spectator stress-current divergence. Its curve component vanishes by diagonality and its spectator component has no \(y\)-dependence. Thus \(\dot P_Y=0\).

Write \(\ell\) for arc length, \(u^m=X'^m/\sqrt{h_{xx}}\) for the unit tangent and \(D_\ell\) for the covariant derivative of \(\gamma\). Varying the weighted length gives the explicit gradient of \(\mathcal T\) and the arc-length term \(\mathcal T u_mD_\ell\delta X^m\). After integration by parts the latter supplies \(D_\ell(\mathcal T u_m)\) to the force. Its tangential part subtracts the tangential component of the explicit gradient. The resulting momentum equation is
\begin{equation}
\dot P_m=\sqrt{h_{xx}}\,\bigl[\mathcal T D_\ell u_m
-(\delta_m{}^n-u_m u^n)\partial_n\mathcal T\bigr].
\label{eq:warped-force}
\end{equation}
Normality preserves \(P_mX'^m=0\), and the spectator equation preserves \(P_Y=0\). Constant flux satisfies Gauss and target-time independence gives \(\dot P_t=0\). The ansatz therefore closes under the full limiting equations. For example, take a Euclidean base with \(a=e^{\lambda_a Z}\) and \(r_Y=e^{\lambda_Y Z}\). A straight curve along \(X^1\) at \(Z=0\) has zero curvature, leaving a purely background force.
\begin{equation}
\dot P_Z=-\lambda_a\sqrt{T_0^2+J^2}
-\frac{\lambda_Y T_0^2}{\sqrt{T_0^2+J^2}} .
\label{eq:warped-straight}
\end{equation}
The first term is due to the gradient of the clock factor, and the second to the gradient of the spectator radius. On the same straight curve with constant magnetic field, Eq.~\eqref{eq:warped-magnetic-force} gives
\begin{equation}
\dot P_Z=-T_0\left(\lambda_a\sqrt{1+B^2}
+\frac{\lambda_Y}{\sqrt{1+B^2}}\right).
\label{eq:warped-magnetic-straight}
\end{equation}
The zero-field and zero-flux forces agree. Both branches couple to the same warp factors, with different gauge variables controlling the local energy density.

\section{Type II Carroll limits and fermionic constraints}
\label{sec:super}

Local \(\kappa\) symmetry has been realized for a flat Carroll superparticle and analyzed in a contraction of a doubled Type II Green--Schwarz superstring \cite{BergshoeffGomisParra2016,OzkanZorba2026}. For D-branes the question is how the two gauge scalings affect the fermionic constraints and their null directions. We take D2 through D7 in flat Type II superspace with constant dilaton, positive orientation and vanishing bosonic Neveu--Schwarz (NS) and Ramond--Ramond (RR) potentials. In the relativistic action the supersymmetric DBI (super-DBI) and Wess--Zumino terms together realize \(\kappa\) symmetry \cite{AganagicPopescuSchwarz1997,BergshoeffTownsend1997}. The auxiliary tensor enters only super-DBI, so the same bosonic reduction provides the starting point for both fermionic calculations.

\subsection{Finite-parameter superbrane time bracket}
\label{sec:finite-super-time}

Eliminate the regular auxiliary pairs at finite parameter. Write \(Z=(X,\theta)\) for the parent superspace embedding, with \(M,N\) target Lorentz-frame indices. The vector component of the supervielbein is \(\Pi^M=\dd X^M-\bar\theta\Gamma^M\dd\theta\), with \(\bar\theta=\theta^{\mathrm T}{\bf C}\) and \({\bf C}=\Gamma_0\) in the real Majorana basis. The induced metric is \(G_{ab}=\eta_{MN}\Pi_a^M\Pi_b^N\), and \(\mathcal F=F-Z^*B_2\) includes the pullback of the target NS two-form. On the worldvolume, \(B_2\) also denotes that pullback. The super-DBI term defines the mechanical momenta \(\widehat P_M,\widehat E^i\). The total canonical momenta \(P_M,\pi^i\) include the Wess--Zumino contributions. The primary fermionic constraint is \(\Psi_{\rm rel}\). In this subsection and the finite bracket calculation of Appendix~\ref{app:super-time-details}, we omit the \({\rm rel}\) label on the unscaled fields and momenta. The canonical one-form separates the mechanical and primary-constraint terms from the Wess--Zumino (WZ) contribution.
\begin{equation}
\vartheta=\int_\Sigma
 \left(\widehat P_M\Pi_\delta^M+\widehat E^i\mathfrak a_i+\Psi_{\rm rel}\,\delta\theta\right)
 +\vartheta_{\rm WZ},\qquad
\mathfrak a_i=\delta A_i-(\iota_{\delta Z}B_2)_i .
\label{eq:super-horizontal-theta}
\end{equation}
Here \(\Pi_\delta\) evaluates the supervielbein on a field variation, and \(\vartheta_{\rm WZ}\) comes from the terms linear in time velocities. The time generator \(H_{\rm rel}^{\rm susy}[f]\) smears half the density \(\widehat P^2+\widehat E^iG_{ij}\widehat E^j+T^2\det m_{\rm susy}\), where \(m_{\rm susy}=(G+\mathcal F)_{\spat}\). Its configuration flow has \(\delta_f\theta=0\), \(\Pi_{\delta_f}^M=f\widehat P^M\) and \(\mathfrak a_i(\delta_f)=fG_{ij}\widehat E^j\). The part of \(\delta\vartheta\) evaluated at fixed \(\widehat P,\widehat E,\Psi_{\rm rel}\) receives contributions from the supervielbein torsion, the NS three-form and the Wess--Zumino term. As shown in Appendix~\ref{app:super-time-details}, spatial integration by parts removes derivatives of the two variations. The two time flows are pointwise proportional, so their contraction with this local antisymmetric curvature vanishes.

Derivatives of the smearings arise only from varying the spatial supervielbein and \(\mathcal F\). Cartan's formula gives \(\dd_\Sigma\mathfrak a-\iota_{\delta Z}\dd B_2\) for \(\delta\mathcal F\). The first term differentiates the smearing, and the supervielbein variation contributes \(\widehat P^M\partial_i f\). The calculation in Sec.~\ref{sec:reduction} now uses \(\Pi_i\) and \(\mathcal F\) in place of \(X_i\) and \(F\). In particular, \(R_{\rm a}G_{\spat}=R_{\rm s}\mathcal F_{\spat}\) for \(R=m_{\rm susy}^{-\mathrm{T}}\). The bracket closes on the mechanical spatial generator \(C_{\rm mech}[v]\), built from the DBI momenta and the supercovariant pullbacks.
\begin{equation}
\begin{gathered}
\{H_{\rm rel}^{\rm susy}[f],H_{\rm rel}^{\rm susy}[g]\}=C_{\rm mech}[v],
\qquad v^j=\mathcal S_{\rm susy}^{ij}b_i,\qquad b_i=f\partial_i g-g\partial_i f,\\
\mathcal S_{\rm susy}^{ij}=\widehat E^i\widehat E^j+T^2\det m_{\rm susy}\,R^{(ij)},\\
C_{\rm mech}[v]=\int_\Sigma v^i
     (\widehat P_M\Pi_i^M+\widehat E^j\mathcal F_{ij}).
\end{gathered}
\label{eq:finite-super-time-bracket}
\end{equation}
Adding the density \(\Psi_{\rm rel}\,\partial_i\theta\) gives the canonical spatial constraint. This follows by evaluating the Wess--Zumino velocity term on a spatial tangent, which gives zero by antisymmetry. Gauss uses \(\pi^i\), the total canonical momentum of \(A_i\).

The canonical Type II analyses give the fermionic bracket and the improvements of the time and spatial generators \cite{HatsudaKamimura1998,KamimuraHatsuda1998}. Appendix~\ref{app:super-time-details} tracks the sixteen null directions and the invertible complementary block through both limits. The primary fermionic flow has zero contraction with \(\Pi_\delta\) and \(\mathfrak a\). Its Cartan variations therefore leave the spinor smearing undifferentiated, making the mixed bracket pointwise proportional to \(f\). Projector derivatives multiply the fermionic constraint. On its surface the Dirac correction pairs identical odd coefficients through a symmetric inverse kernel and vanishes. Thus the improved time generators close weakly on the canonical spatial generator. The explicit correction is evaluated in Appendix~\ref{app:super-time-details}.

Each time generator gains a factor \(c^2\). The parent mechanical flux stays bounded in the magnetic scaling, while \(c\widehat E^i\) stays bounded in the electric scaling. Let \(m_{{\rm susy},c}\) denote the spatial super-DBI block expressed in the rescaled fields. The structure function in either scaling is
\begin{equation}
\mathcal S_{{\rm susy},c}^{ij}
=c^4\widehat E^i\widehat E^j+c^2 T_0^2\det(m_{{\rm susy},c})(m_{{\rm susy},c}^{-1})^{(ij)} .
\label{eq:super-time-structure}
\end{equation}
The structure function is \(O(c^2)\) in both limits. The mechanical spatial density stays bounded since in the electric scaling \(\mathcal F_{ij}=O(c)\) compensates the inverse flux weight. With \(m_{{\rm susy},c}\to h\) and \(c\widehat E^i\to\pi^i\) in that branch, the leading coefficient is
\begin{equation}
\lim_{c\to0}c^{-2}\mathcal S_{{\rm susy},c}^{ij}
=\pi^i\pi^j+T_0^2\det h\,(h^{-1})^{ij}\qquad\text{(electric)} .
\label{eq:super-electric-time-coefficient}
\end{equation}
The magnetic flux square enters at \(O(c^4)\). Both time brackets thus tend to zero at fixed smooth smearings. The limiting actions make this commutativity explicit.

\subsection{Common spinor scaling and Wess--Zumino power counting}

We use the flat Type II supervielbein, field strength and Wess--Zumino forms \cite{AganagicPopescuSchwarz1997,BergshoeffTownsend1997,HatsudaKamimura1998,KamimuraHatsuda1998}. The nonsymmetric-tensor replacement \cite{AbouZeidHull1997} gives the potential in Eq.~\eqref{eq:potential} with the spatial super-DBI block as its second argument. Keep the rescaled fields, their first derivatives and the spatial mechanical momenta bounded. The finite mechanical clock momentum is \(P_t^{\rm mech}=c\widehat P_0^{\rm rel}\), where \(\widehat P_0^{\rm rel}\) is the parent time component of \(\widehat P_M\). In addition to Eq.~\eqref{eq:common-scaling}, take
\begin{equation}
\theta_{\rm rel}=\sqrt c\,\Theta,\qquad
\epsilon_{\rm rel}=\sqrt c\,\epsilon_{\rm C},\qquad
\Pi_{\rm rel}^{0}=c(\dd t-\bar\Theta\Gamma^0\dd\Theta).
\label{eq:super-weights}
\end{equation}
Here \(\epsilon_{\rm C}\) is the rescaled rigid supersymmetry parameter. The spatial supervielbein is \(\Pi_{\rm rel}^I=\dd X^I-c\bar\Theta\Gamma^I\dd\Theta\). The NS two-form starts at fermion degree two, so \(\mathcal F-F=O(c)\) in both scalings.

The exact Wess--Zumino representative \(\Omega_{p+1}^{\rm WZ}\) vanishes at fermion degree zero. Each spinor or spinor differential contributes \(c^{1/2}\), and the tension supplies \(c^{-1}\). The term of degree \(2n\) therefore scales as
\begin{equation}
T_{\rm rel}(\Omega_{p+1}^{\rm WZ})_{[2n]}=O(c^{n-1}),\qquad n\geq1 .
\label{eq:wz-weights}
\end{equation}
An explicit orthonormal time differential supplies another power of \(c\). Thus the possible surviving terms have degree two and use spatial target differentials. The gauge weight decides which of these terms remain.

\subsection{Magnetic Type II limit}
\label{sec:super-full}
\label{sec:super-kappa}

\paragraph{Contracted action.}
With the unscaled gauge pair the spatial block tends to \(m_{\M}=h+F_{\spat}\), whose invertible bosonic part keeps the auxiliary variation regular. Apply the block Legendre transform of Sec.~\ref{sec:parent} to the exact supervelocity and super electric field. In the canonical clock term the factor \(c\) in \(\Pi^0_{\rm rel}\) cancels the inverse power in \(\widehat P_0^{\rm rel}\). This leaves the clock bilinear. The fermionic corrections to the spatial auxiliary potential start at \(O(c)\), while the lapse suppresses the spatial mechanical momentum and electric flux squares.

In the Wess--Zumino term the spatial field strength has order one. The term quadratic in spinors therefore retains the full magnetic polynomial on the spatial target differentials.

For clarity, write \(L=\dd X^I\Gamma_I\), \(L^{[k]}=L^{\wedge k}/k!\) and \(F^{[k]}=F^{\wedge k}/k!\). The matrix-valued forms in the surviving term are
\begin{equation}
\begin{gathered}
\mathcal U_{2r}=\sum_{j=0}^r\Gamma_{11}^{j+1}L^{[2j]}\wedge F^{[r-j]},\\
\mathcal U_{2r+1}=\sum_{j=0}^r L^{[2j+1]}\wedge F^{[r-j]}\tau_3^j\tau_1 .
\end{gathered}
\label{eq:super-polynomial}
\end{equation}
The IIB Pauli matrices act on the common-chirality doublet. The rescaled spinor obeys \(\bar\Theta=\Theta^{\mathrm{T}}{\bf C}\), with \({\bf C}\Gamma^0=\one\). Expand \(\mathcal U_p=\sum_A U_A\omega^A\) in closed ordinary forms. The surviving Wess--Zumino top form is \(-T_0\sum_A(\bar\Theta U_A\dd\Theta)\wedge\omega^A\). Its spatial and gauge velocity coefficients shift the mechanical momenta. Its evaluation with a repeated spatial tangent vanishes, giving the associated shift terms. It has no clock differential, so \(P_t=P_t^{\rm mech}\) in the limit. The canonical density is
\begin{equation}
\begin{gathered}
\mathcal L_{\M}^{\rm susy}
=P_tD_0t+P_ID_0X^I+\pi^i\mathcal E_i
-\Theta^{\mathrm{T}}\boldsymbol D_{\M} D_0\Theta
-\tfrac e2[-P_t^2+T_0^2V(s,m_{\M})],\\
D_0=\partial_0-N^i\partial_i,\qquad
\boldsymbol D_{\M}=P_t\one+T_0\boldsymbol{K},\qquad \boldsymbol{K}={\bf C}[\mathcal U_p]_{\spat}.
\end{gathered}
\label{eq:super-action}
\end{equation}
Here \([\mathcal U_p]_{\spat}\) is the spatial pullback density and \(\mathcal E_i\) is the electric component of \(F\) with the shift contribution subtracted. The spatial auxiliary constraint surface of Sec.~\ref{sec:order-of-reduction} still applies. Its pullback retains the fermionic term controlled by \(\boldsymbol D_{\M}\).

\paragraph{Fermionic constraints.}
The null directions of \(\boldsymbol D_{\M}\) follow from the square of \(\boldsymbol K\). At each point put the spatial field strength in \(r=\lfloor p/2\rfloor\) two-plane blocks in an oriented orthonormal tangent frame, with signed coefficients \(B_\ell\) that may vanish. The zero-flux matrices are
\begin{equation}
\boldsymbol{A}_{2r}=\Gamma_{0\,1\cdots 2r}\Gamma_{11}^{r+1},\qquad
\boldsymbol{A}_{2r+1}=\Gamma_{0\,1\cdots (2r+1)}\tau_3^r\tau_1 .
\end{equation}
The contribution from a plane amounts to deleting its two indices in the Clifford product. In IIA the plane factor is \(\boldsymbol{M}_\ell=\Gamma_{2\ell-1,2\ell}\Gamma_{11}\), and in IIB it is \(\boldsymbol{M}_\ell=-\Gamma_{2\ell-1,2\ell}\tau_3\). Expanding Eq.~\eqref{eq:super-polynomial} in these planes gives
\begin{equation}
\boldsymbol{K}=\sqrt{\det h}\,
\prod_{\ell=1}^{r}(\one+B_\ell\boldsymbol{M}_\ell)\boldsymbol{A}_p .
\label{eq:magnetic-spin-product}
\end{equation}
The plane factors commute with one another and square to minus the identity. Each anticommutes with \(\boldsymbol A_p\), which squares to one. Moving \(\boldsymbol A_p\) through the second product in \(\boldsymbol K^2\) reverses every \(B_\ell\). Each pair gives \(1+B_\ell^2\), so \(\boldsymbol K^2=\Delta^2\one\) with \(\Delta=\sqrt{\det m_{\M}}\). The matrix is symmetric and traceless. Each IIA term anticommutes with chirality, and each IIB term has a traceless doublet factor on the physical spinor space. Hence the two eigenvalues have equal multiplicity.
\begin{equation}
\boldsymbol P_{\M,\pm}=\tfrac12(\one\pm\boldsymbol{K}/\Delta),\qquad
\operatorname{rank}\boldsymbol P_{\M,\pm}=16 .
\label{eq:super-projectors}
\end{equation}
The row momentum \(\zeta\) conjugate to \(\Theta\) obeys \(\Psi_{\M}=\zeta+\Theta^{\mathrm{T}}\boldsymbol D_{\M}=0\). The spatial generator acquires \(\zeta\partial_i\Theta\), and Gauss is unchanged. The bracket of the odd constraints is \(2\boldsymbol D_{\M}\) times the spatial delta function. Appendix~\ref{app:super-time-details} obtains the same bracket by contracting the finite-parameter matrix with \(\Psi_c=\sqrt c\,\Psi_{\rm rel}\) and tracks its rank. The plus block is proportional to the time constraint and supplies sixteen first-class constraints. The minus block is invertible on the future branch and supplies sixteen second-class constraints. Since \(\mathcal C_{\M}\) and \(\boldsymbol D_{\M}\) depend only on \(P_t,X^I,A_i\) and their spatial derivatives, the time generator commutes strongly with itself and with the fermionic constraints. The Dirac correction also vanishes, in agreement with Eq.~\eqref{eq:super-time-structure}.

\paragraph{Local \(\kappa\) symmetry.}
To verify \(\kappa\) invariance of the limiting action, eliminate \(s\) by its regular auxiliary equation. Set \(v=\boldsymbol P_{\M,+}\kappa\), \(\delta\Theta=v\) and \(\delta t=\Theta^{\mathrm T}v\), holding \(X^I\) and \(A_a\) fixed. Variation and integration by parts produce the matrix-density derivative \(\dot{\boldsymbol K}-\partial_i(N^i\boldsymbol K)\). Choose the variations of \(P_I,\pi^i\) to be minus the configuration derivatives of \(T_0\int_\Sigma\Theta^{\mathrm T}\boldsymbol K v\) at fixed \(v\). Their contraction with the shifted velocities cancels that derivative term. The remaining variation of \(S_{\M}^{\rm susy}=\int\dd^{p+1}\sigma\,\mathcal L_{\M}^{\rm susy}\) and the compensating lapse transformation are
\begin{equation}
\begin{gathered}
\delta_\kappa S_{\M}^{\rm susy}=\int\dd^{p+1}\sigma
\left[-2v^{\mathrm{T}}\boldsymbol D_{\M}D_0\Theta-\tfrac12\mathcal C_{\M}\,\delta e\right],\\
\delta e=-\frac{4v^{\mathrm{T}}D_0\Theta}{T_0\Delta-P_t}.
\end{gathered}
\end{equation}
Near the future branch, \(\boldsymbol D_{\M}v=(P_t+T_0\Delta)v=\mathcal C_{\M}v/(T_0\Delta-P_t)\). The lapse variation therefore cancels the remaining term off shell.

The relativistic \(\kappa\)-symmetry matrix \(\Gamma_\kappa\) is a Clifford polynomial divided by the DBI volume factor. The projector on the allowed variations is \((\one+\Gamma_\kappa)/2\) \cite{AganagicPopescuSchwarz1997,BergshoeffTownsend1997}. In our scaling the time component of the pulled-back Clifford form after subtracting the shift contributes \(c v_t\), where \(v_t=-eP_t^{\rm mech}>0\) on the future velocity branch. The shift-subtracted spatial vielbein velocities are \(O(c^2)\) by the Legendre relation. The electric component of \(\mathcal F\) with the shift contribution subtracted is \(e_{\rm rel}G_{ij}\widehat E^j=O(c^2)\) by Eq.~\eqref{eq:velocity} applied to the super-DBI block and the bounded mechanical flux. Thus the leading numerator and denominator are \(c v_t\boldsymbol{K}\) and \(c v_t\Delta\). Their ratio tends to \(\boldsymbol{K}/\Delta\).

\paragraph{Supercharges and independent checks.}
For rigid supersymmetry the variation of the Wess--Zumino form contributes a time-boundary term. Subtracting it from the canonical generator gives the charge with integrand \((\zeta-\Theta^{\mathrm{T}}\boldsymbol D_{\M})\epsilon_{\rm C}\). The opposite sign from that in \(\Psi_{\M}\) makes their mixed bracket vanish. The Dirac correction therefore vanishes, and with \(H_{\M}=-\int_\Sigma P_t\) the algebra is
\begin{equation}
\tfrac12\{Q_{\M},Q_{\M}^{\mathrm{T}}\}_{\rm D}
=H_{\M}\one-T_0\int_\Sigma\boldsymbol{K} .
\label{eq:super-charge}
\end{equation}

The RR Wess--Zumino coupling provides an independent check of the charge matrix through the lower-brane couplings induced by flux \cite{Douglas1995,GreenHullTownsend1996}. For a static D4 or D5 with two orthogonal unit-metric magnetic planes, a factor of \(F\) supplies the corresponding two worldvolume indices. Relative to the parent RR coupling, the four local coefficients are \(1,B_1,B_2,B_1B_2\). In the normalization of Eq.~\eqref{eq:super-charge}, the last coefficient gives the D0 or D1 charge density \(T_0 B_1 B_2\) from \(F\wedge F/2\). Its square equals the mixed term in the energy density squared. On this static solution the local supercharge bracket density is \(T_0(\Delta\one-\boldsymbol{K})\). Its kernel is exactly the image of \(\boldsymbol P_{\M,+}\) and has dimension sixteen. With three magnetic planes on D6/D7, the same argument gives eight coefficients, one for each subset. The sixth-order term \(T_0^2B_1^2B_2^2B_3^2\) is the square of the charge induced by all three planes.

\subsection{Electric Type II limit}
\label{sec:super-electric}

\paragraph{Contracted action.}
Use the electric gauge weights of Eq.~\eqref{eq:escale} with the common clock and spinor weights. Keep the rescaled mechanical flux \(\widehat\pi_c^i=c\widehat E^i\) bounded. In the flat NS convention of the relativistic canonical actions \cite{HatsudaKamimura1998,KamimuraHatsuda1998}, write \(B_2=b_M\wedge(\dd X_{\rm rel}^M-a^M/2)\), with \(a^M=\bar\theta\Gamma^M\dd\theta\) and \(b_M=\bar\theta\Lambda\Gamma_M\dd\theta\). Here \(\Lambda=\Gamma_{11}\) in IIA and \(\Lambda=\tau_3\) in IIB. The time contribution to \(B_2\) and the product of the two currents both start at \(O(c^2)\). Hence
\begin{equation}
\beta_I=\bar\Theta\Lambda\Gamma_I\dd\Theta,\qquad
B_2=c\,\beta_I\wedge\dd X^I+O(c^2),\qquad
\mathcal F_{\rm rel}=c(F-\beta_I\wedge\dd X^I)+O(c^2).
\label{eq:super-electric-ns}
\end{equation}
The spatial block tends to \(h\). With \(W=cw\), the mixed terms are \(w^i\mathfrak E_i/(ce)\) and \(-w^iG_{ij}w^j/(2e)\), where \(\mathfrak E_i=\mathcal F_{{\rm rel},0i}-N^j\mathcal F_{{\rm rel},ji}\). The zero momentum of \(w\) pairs with \(\widehat\pi_c=w/e\) through a mixed bracket of magnitude \(e^{-1}\). The spatial Hessian tends to Eq.~\eqref{eq:hessian} at \(m=h\). Its invertibility and the vanishing auxiliary momenta give the same graph and symplectic pullback in either order. A degree-\(2n\) Wess--Zumino term with \(k\) super field-strength factors scales as \(O(c^{n-1+k})\) including the tension. A target-time leg supplies another power of the contraction parameter. Only the term with \(n=1\), \(k=0\) and spatial target legs survives. A derivative with respect to a parent gauge velocity removes at most one field-strength factor, giving a flux shift of order at most \(O(1)\). Thus \(c\pi_{\rm rel}^i=\widehat\pi_c^i+O(c)\). The extra power from a time leg also suppresses the NS and Wess--Zumino shifts of the rescaled clock momentum. The inverse power in \(\widehat E^i\) keeps the leading NS kinetic term finite. Evaluating the wedge product in Eq.~\eqref{eq:super-electric-ns} gives
\begin{equation}
\lim_{c\to0}\mathfrak E_i/c
=\mathcal E_i-\beta_I(D_0)\partial_iX^I+\beta_{iI}D_0X^I,
\qquad \beta_{iI}=\bar\Theta\Lambda\Gamma_I\partial_i\Theta .
\label{eq:super-electric-velocity}
\end{equation}
The last term adds \(\pi^i\beta_{iI}\) to the Wess--Zumino shift of \(P_I\). The middle term supplies \(\boldsymbol L_\pi={\bf C}\Lambda\Gamma_I J^I\), with \(J^I=\pi^i\partial_iX^I\) and \(J^2=\pi^ih_{ij}\pi^j\). The RR term contributes \(\boldsymbol K_0=\boldsymbol K|_{F_{\spat}=0}=\sqrt{\det h}\,\boldsymbol A_p\). The quartic NS term times the parent flux is \(O(c)\), so the exact limit is quadratic in \(\Theta\).
\begin{equation}
\begin{gathered}
\mathcal L_{\E}^{\rm susy}
=P_tD_0t+P_ID_0X^I+\pi^i\mathcal E_i
-\Theta^{\mathrm T}\boldsymbol D_{\E}D_0\Theta
-\tfrac e2[-P_t^2+J^2+T_0^2V(s,h)],\\
\boldsymbol Z_{\E}=T_0\boldsymbol K_0+\boldsymbol L_\pi,\qquad
\boldsymbol D_{\E}=P_t\one+\boldsymbol Z_{\E}.
\end{gathered}
\label{eq:super-electric-action}
\end{equation}
Eliminating \(s\) gives the electric time constraint \eqref{eq:ce}.

\paragraph{Fermionic constraints.}
The odd momentum gives \(\Psi_{\E}=\zeta+\Theta^{\mathrm T}\boldsymbol D_{\E}\). The spatial constraint acquires \(\zeta\partial_i\Theta\), and Gauss remains \(\partial_i\pi^i\).

To classify the odd constraints, choose an oriented orthonormal spatial tangent frame and let \(\boldsymbol L_{(i)}=\Gamma_0\Lambda\Gamma_i\) in this frame. The electric term squares to \(J^2\one\). In IIA the time and chirality anticommutations give \(\boldsymbol L_{(i)}\boldsymbol A_{2r}=(-1)^{i-1}\Gamma_{1\cdots\widehat i\cdots 2r}\Gamma_{11}^{r}\), with a hat over the omitted index. In IIB the spatial volume product commutes with a tangent gamma matrix, and \(\tau_3\) anticommutes with the RR doublet factor. This gives \(\boldsymbol L_{(i)}\boldsymbol A_{2r+1}=(-1)^{i-1}\Gamma_{1\cdots\widehat i\cdots(2r+1)}\tau_3^{r+1}\tau_1\). In both cases reversing the order changes the sign. Thus
\begin{equation}
\{\boldsymbol L_\pi,\boldsymbol K_0\}_+=0,\qquad
\boldsymbol Z_{\E}^{\,2}=\varepsilon_{\E}^2\one,\qquad
\boldsymbol P_{\E,\pm}=\tfrac12(\one\pm\boldsymbol Z_{\E}/\varepsilon_{\E}) .
\label{eq:super-electric-projectors}
\end{equation}
The zero-flux RR matrix is symmetric by the magnetic calculation. For the electric matrix, \(\Gamma_I^{\mathrm T}{\bf C}=-{\bf C}\Gamma_I\). In IIA transposition through \({\bf C}\) also reverses \(\Gamma_{11}\), and anticommuting it past \(\Gamma_I\) recovers the original product. In IIB \(\tau_3\) is symmetric and commutes with the gamma matrices. Both charge terms are traceless, with the electric term changing sign under conjugation by \(\boldsymbol A_p\). Thus each projector has rank sixteen.

The matrix coefficients depend only on \(P_t,X^I,\pi^i\) and their spatial derivatives, which mutually commute. The odd bracket is \(2\boldsymbol D_{\E}\delta_\Sigma\). Its plus block is \(2\mathcal C_{\E}\boldsymbol P_{\E,+}/(\varepsilon_{\E}-P_t)\), and the minus block is \(-4\varepsilon_{\E}\boldsymbol P_{\E,-}\) on the future branch. Gauss leaves the projectors invariant, and the improved spatial generator transports them as weight-zero matrices. The sixteen first-class and sixteen second-class constraints leave sixteen of the 64 odd phase-space coordinates. The same commuting variables ensure that the time generator commutes strongly with itself and with the odd constraints, so its Dirac time bracket vanishes.

\paragraph{Local \(\kappa\) symmetry.}
To see the local symmetry at the action level, set \(v=\boldsymbol P_{\E,+}\kappa\), \(\delta\Theta=v\) and \(\delta t=\Theta^{\mathrm T}v\). Hold \(X^I,\pi^i,P_t,N^i\) and \(A_0-N^iA_i\) fixed, and take
\begin{equation}
W_v=\int_\Sigma\Theta^{\mathrm T}\boldsymbol Z_{\E}v,\qquad
\delta P_I=-\frac{\delta W_v}{\delta X^I},\qquad
\delta A_i=\frac{\delta W_v}{\delta\pi^i},
\label{eq:super-electric-kappa-variations}
\end{equation}
Take the functional derivatives at fixed \(v,\Theta\). Variation of the clock and odd kinetic terms gives \(-2v^{\mathrm T}\boldsymbol D_{\E}D_0\Theta\) and, after integration by parts, \(\Theta^{\mathrm T}(\partial_0-\mathcal L_N)\boldsymbol Z_{\E}\,v\). The latter is the variation of \(W_v\) along the configuration velocities \(D_0X^I\) and \((\partial_0-\mathcal L_N)\pi^i\). The spatial and gauge kinetic terms contribute \(\delta P_I D_0X^I\) and \(-(\partial_0-\mathcal L_N)\pi^i\,\delta A_i\), respectively, after integration by parts. Equation~\eqref{eq:super-electric-kappa-variations} makes their sum cancel the matrix-density derivative. The remaining variation and lapse transformation are
\begin{equation}
\delta S_{\E}^{\rm susy}
=\int\dd^{p+1}\sigma\left[-2v^{\mathrm T}\boldsymbol D_{\E}D_0\Theta-\tfrac12\mathcal C_{\E}\delta e\right],
\qquad \delta e=-\frac{4v^{\mathrm T}D_0\Theta}{\varepsilon_{\E}-P_t}.
\label{eq:super-electric-kappa-lapse}
\end{equation}
Since \(\boldsymbol D_{\E}v=\mathcal C_{\E}v/(\varepsilon_{\E}-P_t)\), the lapse variation cancels the remainder off shell near the future branch. The extension of this transformation to the action with the auxiliary tensor is given below for both branches.

The parent \(\kappa\) matrix provides an independent check \cite{AganagicPopescuSchwarz1997,BergshoeffTownsend1997}. After subtracting the shift, the time component of the supervielbein is \(cv_t\), with \(v_t=-eP_t=e\varepsilon_{\E}>0\) on the future branch. The spatial vielbein velocities are \(O(c^2)\), while \(\mathfrak E_i=e_{\rm rel}G_{ij}\widehat E^j=c e h_{ij}\pi^j+O(c^2)\). Thus the numerator contains the time component or one electric field-strength component. Each additional spatial field-strength factor supplies another power of \(c\). The two leading terms are \(cv_t\boldsymbol K_0\) and \(cv_t(\boldsymbol L_\pi/\varepsilon_{\E})\boldsymbol K_0\), with the relative sign fixed by the Clifford products with the corresponding indices omitted. The leading DBI volume is \(cv_t\sqrt{\det h}\sqrt{1-J^2/\varepsilon_{\E}^2}\). Consequently, with \(\ell=\boldsymbol L_\pi/\varepsilon_{\E}\) and \(\lambda=T_0\sqrt{\det h}/\varepsilon_{\E}\),
\begin{equation}
\lim_{c\to0}\Gamma_\kappa=(\one+\ell)\boldsymbol A_p/\lambda,\qquad
\lambda^2=1-J^2/\varepsilon_{\E}^2 .
\label{eq:super-electric-kappa-parent}
\end{equation}
The canonical plus eigenvalue equation is \((\one-\ell)v=\lambda\boldsymbol A_p v\). Multiplying it by \(\one+\ell\) gives \((\one+\ell)\boldsymbol A_p v=\lambda v\), and the implication reverses because \(\lambda>0\). The two projectors have the same image.

\paragraph{Supercharges and independent checks.}
For a constant \(\epsilon_{\rm C}\), the supersymmetry variation of the NS two-form induces a shift of \(A_i\) by \(-\Theta^{\mathrm T}{\bf C}\Lambda\Gamma_I\partial_iX^I\epsilon_{\rm C}\), accompanying \(\delta\Theta=\epsilon_{\rm C}\) and \(\delta t=-\Theta^{\mathrm T}\epsilon_{\rm C}\). The canonical generator with the boundary term from the Wess--Zumino variation subtracted is
\begin{equation}
Q_{\E}[\epsilon_{\rm C}]=\int_\Sigma(\zeta-\Theta^{\mathrm T}\boldsymbol D_{\E})\epsilon_{\rm C},\qquad
\tfrac12\{Q_{\E},Q_{\E}^{\mathrm T}\}_{\rm D}=
-\int_\Sigma P_t\,\one-\int_\Sigma\boldsymbol Z_{\E}.
\label{eq:super-electric-charge}
\end{equation}
The sign in the generator makes its mixed bracket with \(\Psi_{\E}\) vanish. Its canonical self-bracket therefore gives the Dirac algebra above. The vector \(J^I\) is the local fundamental-string charge density along the electric displacement, as expected from the relativistic coupling to the NS two-form \cite{DiVecchiaFrauLerdaLiccardo2000}. For a single-valued embedding coordinate, integrating by parts reduces its charge to a boundary term on the Gauss surface. Winding or compatible boundary data can support a nonzero integrated charge.

As a direct DBI test, set the spinor to zero on a plane with constant \(h\) and constant flux. Take \(\mathcal F_{{\rm rel},0i}=c u_i\) and zero shift and spatial field strength. The Schur complement of \(h\) gives
\begin{equation}
-\det M=c^2\det h\,(v_t^2-u^{\mathrm T}h^{-1}u),\qquad
\pi=\frac{T_0\sqrt{\det h}\,h^{-1}u}{\sqrt{v_t^2-u^{\mathrm T}h^{-1}u}} .
\label{eq:super-electric-planar}
\end{equation}
Differentiation with respect to \(v_t\) gives \(P_t=-T_0\sqrt{\det h}\,v_t/\sqrt{v_t^2-u^{\mathrm T}h^{-1}u}\). Inserting \(u=v_t h\pi/\varepsilon_{\E}\) gives \(P_t=-\varepsilon_{\E}\) and \(\sqrt{-\det M}=cv_t T_0\det h/\varepsilon_{\E}>0\). This tests the positive DBI branch and the normalization of Eq.~\eqref{eq:super-electric-kappa-parent} even with a nonorthogonal spatial metric.

\subsection{Dirac stabilization and the auxiliary \texorpdfstring{\(\kappa\)}{kappa} transformation}
\label{sec:super-stabilization}

In both actions the variation of the auxiliary block's argument \(m\) vanishes under the corresponding \(\kappa\) transformation. The magnetic variation fixes both \(X^I\) and \(A_a\), so it preserves \(m=h+F_{\spat}\). The electric variation fixes \(X^I\) and hence \(m=h\). The nondegenerate auxiliary Hessian then permits local coordinates at fixed \(m\) in which \(V(s,m)-\det m\) is quadratic. Varying each such coordinate by minus one half the relative lapse variation times that coordinate cancels the lapse variation of the additional term in the action. This extends the reduced transformations to both actions with \(s\) retained.

The even stabilization of Sec.~\ref{sec:dirac-completeness} also applies to the Type II parent and the two limiting actions. In the parent mixed pair, the flux relation uses the mechanical momentum, so its canonical form is \(\pi^i-\pi_{\rm WZ}^i-w^i/e=0\), with \(\pi_{\rm WZ}^i\) the Wess--Zumino shift in the same scaling as \(\pi^i\). The shift is independent of \(w\) and of all multipliers. The mixed bracket with \(p_{w,i}\) is therefore the one in Eq.~\eqref{eq:mixed-primary-bracket}. Any brackets among the flux relations leave the mixed block invertible. The mixed pair may couple to the fermionic primaries, so their brackets are evaluated after pulling back to the mixed constraint surface. The spatial Hessian is unchanged in form, with the finite super-DBI block as its argument. The inverse exists when the bosonic body is invertible. The auxiliary pullbacks then leave the canonical odd momentum and its primary constraint.

After auxiliary reduction, write \(\Psi_\pm\) for independent local components of the first- and second-class fermionic sets, with \(C_-=\{\Psi_-,\Psi_-\}\) invertible on the constraint surface. Use the stabilized time generator from Appendix~\ref{app:super-time-details}, the Gauss-improved spatial generator and Gauss to form the bosonic-constraint part \(H_{\rm ev}\) of the total Hamiltonian. These representatives differ from the mechanical ones by constraints. Let \(u_\pm\) be the odd multipliers, with the ordering chosen so that the multiplier term in the preservation equation for \(\Psi_-\) is \(C_-u_-\). Then
\begin{equation}
\dot\Psi_-\simeq\{\Psi_-,H_{\rm ev}\}+C_-u_-,
\qquad
u_-\simeq-C_-^{-1}\{\Psi_-,H_{\rm ev}\},
\qquad
\dot\Psi_+\simeq0.
\label{eq:odd-stabilization}
\end{equation}
In the finite theory the improved even constraints and the independent \(\kappa\) generators are first class by the Type II stabilization identities \cite{HatsudaKamimura1998,KamimuraHatsuda1998}. In both Carroll limits the time and Gauss brackets with the odd constraints vanish strongly, and the improved spatial flow acts by the Lie derivative. The plus projection has a bracket proportional to the time constraint in each branch. These brackets give the last equality in Eq.~\eqref{eq:odd-stabilization} and leave \(u_+\) arbitrary. The even secondary constraints are preserved after imposing the odd second-class set using its Dirac bracket. Equation~\eqref{eq:first-class-lift} gives equivalent first-class representatives with the second-class set retained. Thus there is no additional odd or even constraint on this patch. The independent even generators have the same count on the bosonic body, so turning on the nilpotent fermionic terms preserves their local independence. With 32 spinor components there are 64 odd phase-space coordinates. The sixteen first-class and sixteen second-class constraints leave sixteen odd coordinates, alongside the eight even canonical pairs. This classification is local and assumes the regular rank conditions stated above.

\section{Spatial T-duality of the two Carroll limits}
\label{sec:spatial-duality}

\subsection{Common relativistic map on circle zero modes}

For both contractions, use double dimensional reduction on the wrapped side. Fix the embedding along a flat circle of physical circumference \(L_z\) with unit positive winding, and keep fields and gauge parameters independent of the circle coordinate. In the D2/D3 pair the circle lies along the IIB D3 and is transverse to the IIA D2. Both targets are ten-dimensional, and the shared base has nine dimensions including time. For integrated charges we assume periodic spinors and compatible flux and boundary data. The finite D-brane map exchanges the wrapped gauge component with a transverse IIA scalar \(Y\) \cite{BergshoeffdeRoo1996,KamimuraSimon2000}. In our orientation it takes the form
\begin{equation}
A_{{\rm B},z}=-Y,\qquad T_{\rm A}=L_zT_{\rm B},\qquad
\theta_{\rm B}=S\theta_{\rm A}=(\Gamma_z\theta_-,-\theta_+).
\label{eq:super-tmap}
\end{equation}
The labels \({\rm A},{\rm B}\) refer to IIA and IIB, respectively. The IIA components obey \(\Gamma_{11}\theta_\pm=\pm\theta_\pm\). The constant isometry \(S\) maps them to an IIB doublet of positive chirality. The chirality flip of one component is the flat-circle instance of the Type II spinor transformation \cite{Hassan2000}. The radii satisfy \(R_{\rm A}R_{\rm B}=\alpha'\) with \(L_z=2\pi R_{\rm B}\). The dilaton shift gives \(g_{s,\rm A}=g_{s,\rm B}\sqrt{\alpha'}/R_{\rm B}\) \cite{Buscher1988,KamimuraSimon2000}. With \(T_p=[(2\pi)^p g_s(\alpha')^{(p+1)/2}]^{-1}\), the tension ratio is \(L_z\) before and after rescaling by the common \(c\). In the exact supergeometry the transverse and wrapped one-forms are \(u_y=\Pi_{\rm A}^y\) and \(k_z=\Pi_{\rm B}^z\). Their field-strength relation is
\begin{equation}
\mathcal F_{\rm B}=\mathcal F_{\rm A}-u_y\wedge k_z,\qquad u_y=\Pi_{\rm A}^y .
\end{equation}
The change from \(\dd z\) to \(k_z\) has unit determinant. In that coframe, the wrapped DBI matrix and its circle Schur complement are
\begin{equation}
M_{\rm B}=
\begin{pmatrix}
G_{\parallel}+\mathcal F_{\rm A} & -u_y\\
u_y^{\mathrm{T}}&1
\end{pmatrix},
\qquad
M_{\rm A}=G_{\parallel}+\mathcal F_{\rm A}+u_yu_y^{\mathrm{T}},
\end{equation}
where \(G_{\parallel}\) is the common induced metric with the circle omitted. Thus both full and spatial DBI determinants match at finite parameter.

The relativistic map also relates the RR Wess--Zumino couplings \cite{GreenHullTownsend1996,KamimuraSimon2000}. To check the curvature in our orientation, evaluate the polynomial in Eq.~\eqref{eq:super-polynomial} on the exact superforms between two fermion differentials. The circle component enters through a Clifford one-form or the term \(-u_y\wedge k_z\) in the super field strength. With \(\Gamma_{[n]}\) a product of \(n\) Clifford matrices along the shared directions, the spinor identities are
\begin{equation}
\begin{gathered}
S^{\mathrm{T}}({\bf C}\Gamma_{[2j]}\Gamma_z\tau_3^j\tau_1)S
={\bf C}\Gamma_{[2j]}\Gamma_{11}^{j+1},\\
S^{\mathrm{T}}({\bf C}\Gamma_{[2j+1]}\tau_3^j\tau_1)S
=-{\bf C}\Gamma_{[2j+1]}\Gamma_z\Gamma_{11}^{j}.
\end{gathered}
\label{eq:super-tcurvature}
\end{equation}
The second map cancels the minus sign from \(-u_y\wedge k_z\) and restores the transverse Clifford term. The curvatures match after circle integration. Using the same radial homotopy in the fermionic coordinates and a vanishing bosonic representative gives matching local potentials with the tension factor in Eq.~\eqref{eq:super-tmap}. An exact change of representative shifts the canonical one-form by a field-space exact term under the stated boundary conditions and changes the Noether surface term accordingly. The finite-parameter Legendre map thus identifies the symplectic forms and improved charges. The mechanical circle constraint fixes \(\widehat P_{{\rm B},z}=-u_{y,i}\widehat E_{\rm B}^{\,i}\). The transverse momentum on the IIA side is \(-L_z(\widehat E_{\rm B}^{\,z}+k_{z,i}\widehat E_{\rm B}^{\,i})\). Their squares account, respectively, for the transverse gradient in the IIA electric flux square and for the IIB wrapped electric flux square. With the spatial determinant identity this gives \(\mathcal C_{{\rm A},{\rm rel}}=L_z^2\mathcal C_{{\rm B},{\rm rel}}\) on the circle constraint.

\subsection{Magnetic circle duality}
\label{sec:super-tduality}

\paragraph{Duality before contraction.}
The finite map preserves the magnetic weights of the circle pair. The mechanical circle variables stay finite, with \(k_{z,i}=O(c)\), so their explicit momentum squares are suppressed by the rescaled lapse. The transverse gradient still contributes to the spatial DBI block through the Schur complement. Taking the limit in the common finite determinant, symplectic form and Wess--Zumino relations therefore gives the magnetic Hamiltonian and charge matrices on each side.

\paragraph{Contraction before duality.}

Fix the wrapped coordinate in Eq.~\eqref{eq:super-action} and solve its spatial constraint. Its embedding term drops out of the pulled-back canonical potential. The gauge term \(\pi_{\rm B}^{\,z}\delta A_{\rm B,z}=-\pi_{\rm B}^{\,z}\delta Y\) then gives the transverse momentum after circle integration. The spinor isometry gives the fermionic map. The lapse transforms inversely with the circumference because the time constraint scales quadratically.
\begin{equation}
P_{{\rm A},Y}=-L_z\pi_{\rm B}^{\,z},\qquad
\zeta_{\rm A}=L_z\zeta_{\rm B}S,\qquad e_{\rm A}=e_{\rm B}/L_z .
\label{eq:super-tphase}
\end{equation}
Each shared total momentum is multiplied by \(L_z\). The term \(L_z\pi_{\rm B}^{\,z}F_{iz}\) gives the transverse scalar term in the spatial generator. The circle derivative drops out of Gauss, and the time constraint rescales by \(L_z^2\).

\paragraph{Fermionic and supercharge maps.}
In the limiting theory the same spinor identities in Eq.~\eqref{eq:super-tcurvature} apply to ordinary spatial differentials. This identifies the charge matrices, the determinant factors and the improved charges for a common parameter \(\epsilon_{\rm C}\).
\begin{equation}
\boldsymbol{K}_{\rm A}=S^{\mathrm{T}}\boldsymbol{K}_{\rm B}S,\qquad
\Delta_{\rm A}=\Delta_{\rm B},\qquad
Q_{\rm A}[\epsilon_{\rm C}]=Q_{\rm B}[S\epsilon_{\rm C}].
\label{eq:super-tcharge}
\end{equation}
The two orders give the same symplectic pullback and constraint map on the regular, gauge-fixed zero modes of D2/D3, D4/D5 and D6/D7. The matrix identity intertwines the fermionic blocks and the Dirac supercharge algebra. It also identifies the limiting Lagrangian \(\kappa\) projectors because their magnetic matrices are \(\boldsymbol K/\Delta\).

For two magnetic planes in the D5 base, the induced D1 wraps the dual circle and maps to the D4 D0 charge. A plane containing the circle tests a different component of the map. On the positively oriented D5 take
\begin{equation}
\begin{gathered}
F_{\rm B}=B\,\dd\sigma^1\wedge\dd\sigma^2+b\,\dd\sigma^3\wedge\dd z,\\
F_{\rm A}=B\,\dd\sigma^1\wedge\dd\sigma^2,\qquad \dd Y=-b\,\dd\sigma^3 .
\end{gathered}
\label{eq:circle-flux-example}
\end{equation}
The D4 scalar slope changes its metric entry in direction three to \(1+b^2\). This reproduces the factor from the second magnetic plane in the D5 determinant. In the orientation with the wrapped direction last, the term \(F_{\rm B}\wedge F_{\rm B}/2\) gives the signed D1 probe coefficient \(-Bb\) along direction four in units of \(T_{{\rm B},0}\). On the D4, \(F_{\rm A}\) supplies the first plane and \(\dd Y=-b\,\dd\sigma^3\) supplies the circle leg of the tilted D2 charge. Its oriented \(4,y\) component is \(+Bb\) in units of \(T_{{\rm A},0}\). The D4 D0 charge vanishes because \(F_{\rm A}\wedge F_{\rm A}=0\). The mixed energy-square term is now the square of the D2 component with the circle leg. Equation~\eqref{eq:super-tcharge} identifies the sixteen preserved spinors on the two sides.

\subsection{Electric circle duality and rescaling of the transverse pair}
\label{sec:super-electric-tduality}

Use the same fixed physical radii and circle zero modes. The finite map \eqref{eq:super-tmap}, applied to the rescaled electric pair on the wrapped brane, gives \(Y=-cA_{{\rm B},z}\) and \(P_{{\rm A},Y}^{\rm rel}=-L_z\pi_{\rm B}^z/c\). Thus the transverse pair on the IIA side requires its own weights. Put
\begin{equation}
Y=c\mathcal Y,\qquad p_{\mathcal Y}=cP_{{\rm A},Y}^{\rm rel},\qquad
\mathcal Y=-A_{{\rm B},z},\qquad p_{\mathcal Y}=-L_z\pi_{\rm B}^z .
\label{eq:super-electric-tdual-weights}
\end{equation}
We call Eq.~\eqref{eq:super-electric-tdual-weights} the adapted transverse scaling because the finite T-duality map fixes the relative powers of the coordinate and momentum. The physical radius is fixed. The period in \(\mathcal Y\) is \(2\pi R_{\rm A}/c\). The local comparison uses bounded rescaled fields and gradients, including nonconstant \(A_{{\rm B},z}\) and nonzero \(\pi_{\rm B}^z\). For integrated charges we impose the same compatible boundary data on both sides. A fixed nonzero transverse winding that forces unbounded gradients in \(\mathcal Y\) lies outside this domain.

\paragraph{Duality before contraction.}
Apply the finite circle map and then take the electric limit of the IIA action with Eq.~\eqref{eq:super-electric-tdual-weights}. The transverse vielbein is \(\Pi_{\rm A}^y=c(\dd\mathcal Y-\bar\Theta_{\rm A}\Gamma^y\dd\Theta_{\rm A})\). Like the time component, it is of order \(c\), so \(u_{y,i}=O(c)\) and \(k_{z,i}=O(c)\) in the finite Schur calculation. Both super field strengths are \(O(c)\). The spatial IIA and wrapped IIB determinants tend to \(\det h\), with \(h\) now the induced metric of the shared spatial embeddings. Their limiting spatial blocks are \(h\) and \(\operatorname{diag}(h,1)\), respectively. The mixed and spatial auxiliary pairs are regular by Secs.~\ref{sec:dirac-completeness} and \ref{sec:super-stabilization}, since the mixed bracket has magnitude \(e^{-1}\) and the spatial Hessian has invertible body in each case.

The exact mechanical relation is
\begin{equation}
c\widehat P_{{\rm A},Y}
=-L_z\bigl(c\widehat E_{\rm B}^z+k_{z,i}\,c\widehat E_{\rm B}^i\bigr)
\ \longrightarrow\ -L_z\pi_{\rm B}^z .
\label{eq:electric-circle-mechanical}
\end{equation}
The term containing \(k_{z,i}\) vanishes. The parent Wess--Zumino flux shift is at most \(O(1)\), as in Sec.~\ref{sec:super-electric}, so its rescaled contribution also vanishes. On the IIA side the transverse NS term is \(O(c^2)\) from the spinor pair and the transverse differential. Its product with the parent flux is \(O(c)\). The transverse Wess--Zumino velocity term is suppressed by the same additional differential power. The surviving NS flux term therefore uses only shared spatial indices, while the transverse vielbein retains its own fermion bilinear in the canonical kinetic term. With \(\boldsymbol K_{{\rm A},0}\) the spatial RR matrix at zero magnetic field, the new charge matrix and exact limiting action density are
\begin{equation}
\begin{aligned}
\boldsymbol L_{{\rm A},\parallel}&={\bf C}\Gamma_{11}\Gamma_I\,\pi_{\rm A}^i\partial_iX^I,
&
\boldsymbol D_{\rm A}&=P_{{\rm A},t}\one+\boldsymbol Z_{\rm A},\\
\boldsymbol Z_{\rm A}
&=T_{{\rm A},0}\boldsymbol K_{{\rm A},0}
+\boldsymbol L_{{\rm A},\parallel}+p_{\mathcal Y}{\bf C}\Gamma_y,\\
\mathcal L_{\rm A}
&=P_{{\rm A},t}D_0t+P_{{\rm A},I}D_0X^I+p_{\mathcal Y}D_0\mathcal Y
+\pi_{\rm A}^i\mathcal E_i-\Theta_{\rm A}^{\mathrm T}\boldsymbol D_{\rm A}D_0\Theta_{\rm A}\\
&\quad-\tfrac{e_{\rm A}}2\bigl[-P_{{\rm A},t}^2+p_{\mathcal Y}^2
+\pi_{\rm A}^ih_{ij}\pi_{\rm A}^j+T_{{\rm A},0}^2V(s,h)\bigr].
\end{aligned}
\label{eq:electric-circle-action}
\end{equation}
Here \(I\) excludes the transverse circle direction. The shifted electric field of the ordinary gauge potential is \(\mathcal E_i\), and the momenta without hats are total canonical momenta. At \(s=h\) the potential is \(\det h\). The term \(p_{\mathcal Y}^2\) is the limit of \(c^2\widehat P_{{\rm A},Y}^2\). The primary fermionic relation is \(\Psi_{\rm A}=\zeta_{\rm A}+\Theta_{\rm A}^{\mathrm T}\boldsymbol D_{\rm A}\). Multiplying the finite mechanical time-density relation by \(c^2\) gives the same limiting clock, transverse and base flux squares with the common circumference factor.

\paragraph{Contraction before duality.}
Start with the wrapped IIB electric action after contraction. Pair the wrapping condition \(X_{\rm B}^z=z\) with \(\mathcal C_{{\rm B},z}=P_{{\rm B},z}-\pi_{\rm B}^i\partial_i A_{{\rm B},z}\). Their mixed bracket is the unit spatial delta. Solving this pair removes \(P_{{\rm B},z}\delta X_{\rm B}^z\) from the symplectic pullback. The integrated circle term \(L_z\pi_{\rm B}^z\delta A_{{\rm B},z}\) then becomes \(p_{\mathcal Y}\delta\mathcal Y\). Each shared momentum is multiplied by \(L_z\), with \(\Theta_{\rm B}=S\Theta_{\rm A}\) and \(\zeta_{\rm A}=L_z\zeta_{\rm B}S\). The resulting canonical one-form agrees with the one used in Eq.~\eqref{eq:electric-circle-action}.

The IIB electric square splits into the base flux square and \((\pi_{\rm B}^z)^2\). The spatial determinant is \(\det h\). With \(T_{{\rm A},0}=L_zT_{{\rm B},0}\), the time densities therefore satisfy \(\mathcal C_{\rm A}=L_z^2\mathcal C_{\rm B}\), and the lapse is \(e_{\rm A}=e_{\rm B}/L_z\). The spatial term \(L_z\pi_{\rm B}^zF_{{\rm B},iz}=p_{\mathcal Y}\partial_i\mathcal Y\) supplies the scalar contribution in \(\mathcal C_{{\rm A},i}=L_z\mathcal C_{{\rm B},i}\). The circle derivative drops from Gauss, so \(\mathcal G_{\rm A}=L_z\mathcal G_{\rm B}\). Thus the two orders induce the same symplectic and bosonic constraint maps. At the adapted endpoint the time density and the fermionic matrix use mutually commuting variables \(P_{{\rm A},t},p_{\mathcal Y},X^I,\pi_{\rm A}^i\) and their spatial derivatives. The time generator commutes strongly with itself and the odd constraints.

\paragraph{Fermionic and supercharge maps.}
On the wrapped slice, the electric IIB matrix contains the shared NS term and \(\pi_{\rm B}^z{\bf C}\tau_3\Gamma_z\) in addition to its zero-magnetic-flux RR term. The finite spinor isometry gives \cite{KamimuraSimon2000}
\begin{equation}
\begin{gathered}
S^{\mathrm T}({\bf C}\tau_3\Gamma_I)S={\bf C}\Gamma_{11}\Gamma_I,\qquad
S^{\mathrm T}({\bf C}\tau_3\Gamma_z)S=-{\bf C}\Gamma_y,\\
S^{\mathrm T}\boldsymbol K_{{\rm B},0}S=\boldsymbol K_{{\rm A},0},\qquad
\boldsymbol D_{\rm A}=L_z S^{\mathrm T}\boldsymbol D_{\rm B}S .
\end{gathered}
\label{eq:electric-circle-spinor}
\end{equation}
The RR identity is the zero-magnetic-flux term of Eq.~\eqref{eq:super-tcurvature}; the second NS identity fixes the sign of the transverse momentum term independently of the bosonic energy square. It follows that \(\Psi_{\rm A}=L_z\Psi_{\rm B}S\). Since the future energy densities satisfy \(\varepsilon_{\rm A}=L_z\varepsilon_{\rm B}\), the normalized plus and minus projectors are intertwined by \(S\). Thus the map identifies the sixteen local \(\kappa\) directions and the complementary sixteen second-class directions. As a check on the IIA Clifford square, \({\bf C}\Gamma_y\) squares to the identity and anticommutes with the shared flux matrix and the zero-magnetic-flux RR matrix. This gives \(\boldsymbol Z_{\rm A}^2=\varepsilon_{\rm A}^2\one\) with the energy of Eq.~\eqref{eq:electric-circle-action}.

The Lagrangian \(\kappa\)-symmetry matrices also agree under the spinor map. In the adapted IIA velocity frame the leading time and transverse supervielbein components are \(cv_t\) and \(cv_y\), with \(v_t=e_{\rm A}\varepsilon_{\rm A}\) and \(v_y=e_{\rm A}p_{\mathcal Y}\). The shift-subtracted electric component of the supercovariant field strength has leading term \(cu_i\), with \(u_i=e_{\rm A}h_{ij}\pi_{\rm A}^j\). The DBI volume at this order is \(c\sqrt{\det h}\sqrt{v_t^2-v_y^2-u^{\mathrm T}h^{-1}u}\). The time component of the pulled-back Clifford form gives the zero-field brane matrix \(\boldsymbol K_{{\rm A},0}\), one electric field-strength component gives the shared NS matrix times that brane matrix, and the transverse component of the Clifford form gives \({\bf C}\Gamma_y\boldsymbol K_{{\rm A},0}\) times its velocity. Define
\(\ell_{\rm A}=(\boldsymbol L_{{\rm A},\parallel}+p_{\mathcal Y}{\bf C}\Gamma_y)/\varepsilon_{\rm A}\)
and \(\lambda_{\rm A}=T_{{\rm A},0}\sqrt{\det h}/\varepsilon_{\rm A}\). The normalized projector matrix is
\begin{equation}
\Gamma_{\kappa,{\rm A}}\ \longrightarrow\
(\one+\ell_{\rm A})\boldsymbol A_p/\lambda_{\rm A},\qquad
\lambda_{\rm A}^2=1-\frac{p_{\mathcal Y}^2+\pi_{\rm A}^ih_{ij}\pi_{\rm A}^j}{\varepsilon_{\rm A}^2}.
\label{eq:electric-circle-kappa}
\end{equation}
Here \(p\) is the spatial dimension of the IIA brane. Equation~\eqref{eq:electric-circle-spinor} identifies \(\ell_{\rm A}\) and \(\boldsymbol A_p\) with their wrapped counterparts, and \(\lambda_{\rm A}=\lambda_{\rm B}\). The canonical plus eigenspace obeys \((\one-\ell_{\rm A})v=\lambda_{\rm A}\boldsymbol A_pv\); multiplying by \(\one+\ell_{\rm A}\) gives the positive eigenspace of the Lagrangian \(\kappa\)-symmetry matrix.

Finally, the phase-space generator of rigid supersymmetry has integrand \((\zeta_{\rm A}-\Theta_{\rm A}^{\mathrm T}\boldsymbol D_{\rm A})\epsilon_{\rm C}\). In particular the transverse term generates the scaled scalar shift \(-\Theta_{\rm A}^{\mathrm T}{\bf C}\Gamma_y\epsilon_{\rm C}\). The Wess--Zumino boundary contribution to the charge maps with the chosen finite potentials. Its continuation to the endpoint and Eq.~\eqref{eq:electric-circle-spinor} give
\begin{equation}
Q_{\rm A}[\epsilon_{\rm C}]=Q_{\rm B}[S\epsilon_{\rm C}],\qquad
\{Q_{\rm A},Q_{\rm A}^{\mathrm T}\}_{\rm D}
=S^{\mathrm T}\{Q_{\rm B},Q_{\rm B}^{\mathrm T}\}_{\rm D}S .
\label{eq:electric-circle-charge}
\end{equation}
The matching symplectic forms and second-class blocks imply equality of the Dirac algebras under the map. Thus for D2/D3, D4/D5 and D6/D7 the fixed-radius electric zero-mode comparison gives the same reduced phase space and supercharges in both orders, with the adapted transverse scaling on the dual side. Setting the spinors to zero gives the bosonic maps. The gradient of the dual scalar survives in the magnetic determinant, and its momentum survives in the electric constraint.

\section{Carroll--Weyl transformations and global dilations}
\label{sec:carroll-weyl}

The tensionless string admits a geometric action with a worldsheet vector density \cite{IsbergLindstromSundborgTheodoridis1994}. A gauged null-string formulation makes the local temporal and spatial rescalings of the Carrollian worldsheet geometry explicit \cite{SheikhJabbariYavartanoo2026}. Here we test a specific matter transformation, following the canonical construction for non-BPS D-branes in Ref.~\cite{ZengPhase2026}. We rescale the spatial embedding and gauge potential with a worldvolume-dependent parameter, at fixed positive \(T_0\) in the flat single-clock theory. The relativistic brane examples with a compensating scalar motivate testing whether the existing auxiliary tensor can compensate the matter variation \cite{GarciaLinaresVergara2001}. We impose its stationary equation and perform this test in both branches. We then derive the constant dilations that also act on the clock pair.

\subsection{Canonical matter transformation and its obstructions}

In this section \(\chi(\sigma,\tau)\) denotes the external Weyl parameter. The corresponding charge on our matter phase space and its action on a function \(\mathcal O\) are
\begin{equation}
Q_\chi=\int_\Sigma\chi(P_I X^I+\pi^i A_i),\qquad
\delta_\chi\mathcal O=\{\mathcal O,Q_\chi\}.
\label{eq:cw-charge}
\end{equation}
The canonical brackets give weights \(+1\) for \(X^I,A_i\) and \(-1\) for \(P_I,\pi^i\). The clock pair is inert. At fixed \(\chi\), the variation of each product in the equal-time symplectic potential cancels between the two partners. Thus \(Q_\chi\) generates a symplectic transformation for any prescribed smearing.

Spatial differentiation introduces gradients of \(\chi\). With \(u_i=X_I\partial_i X^I\), the variations of the induced metric and of the field strength are
\begin{equation}
\begin{gathered}
\delta_\chi h_{ij}=2\chi h_{ij}+u_i\partial_j\chi+u_j\partial_i\chi,\\
\delta_\chi F_{ij}=\chi F_{ij}+(\partial_i\chi)A_j-(\partial_j\chi)A_i .
\end{gathered}
\label{eq:cw-config}
\end{equation}
Using the covariant spatial constraint in Sec.~\ref{sec:magnetic}, the corresponding variations of the Gauss and spatial densities are
\begin{equation}
\begin{gathered}
\delta_\chi\mathcal G=-\chi\mathcal G-\pi^i\partial_i\chi,\\
\delta_\chi\mathcal C_i
=(P_I X^I+\pi^j A_j)\partial_i\chi-A_i\pi^j\partial_j\chi .
\end{gathered}
\label{eq:cw-gauss-spatial}
\end{equation}
The homogeneous terms cancel between each momentum and its corresponding spatial gradient. The two terms containing an undifferentiated potential come from \(\pi^j\delta_\chi F_{ij}\). Passing to the Gauss-improved spatial generator preserves the tangency criterion for the pair of spatial and Gauss constraints.

For constant \(\chi\), the clock square is invariant and the metric weight of the electric flux square cancels the two flux weights. The determinant variation \(\delta\det m=\det m\,\tr(m^{-1}\delta m)\) gives
\begin{equation}
\begin{gathered}
\delta_\chi\mathcal C_{\E}=2p\chi T_0^2\det h,\\
\delta_\chi\mathcal C_{\M}=\chi T_0^2\det(h+F_{\spat})
  \tr\!\big[(h+F_{\spat})^{-1}(2h+F_{\spat})\big].
\end{gathered}
\label{eq:cw-hamiltonian}
\end{equation}
The electric coefficient is positive on the regular domain. To examine the magnetic coefficient, write \(B_\ell\) for the signed components of \(F\) in its two-plane blocks in an \(h\)-orthonormal frame. The scaled determinant contains a factor \(e^{4\chi}+e^{2\chi}B_\ell^2\) in each plane and \(e^{2\chi}\) in each unpaired direction. Its logarithmic derivative gives the magnetic trace.
\begin{equation}
2p-2\sum_{\ell=1}^{\lfloor p/2\rfloor}
            \frac{B_\ell^2}{1+B_\ell^2}>0.
\label{eq:cw-magnetic-positive}
\end{equation}

The obstruction also occurs on the full constraint surface. On a flat patch take \(X^I=\sigma^I\) in the tangent directions and constant \(t\). Choose \(F_{12}=B\) and \(\pi^1=J\), with other flux components zero. Set \(P_I\partial_i X^I=-F_{ij}\pi^j\). Gauss and the spatial constraints vanish, and the future clock roots are
\begin{equation}
P_{t,\M}=-T_0\sqrt{1+B^2},\qquad
P_{t,\E}=-\sqrt{T_0^2+J^2}.
\end{equation}
On these data Eq.~\eqref{eq:cw-hamiltonian} gives \(2\chi T_0^2[p+(p-1)B^2]\) and \(2p\chi T_0^2\), respectively. Set \(A_i=0\) by a gauge choice at the origin, where \(X^I=0\). This removes the gradient terms in Eq.~\eqref{eq:cw-config} at that point and leaves the algebraic obstruction for local profiles with nonzero value there. First-class improvements have weakly vanishing brackets with the constraints, and multiplier variations multiply the constraints themselves. The on-shell coefficients are therefore unchanged by these choices.

The action with the auxiliary tensor retained gives the same test on its auxiliary constraint surface. The block \(m\) equals \(m_{\M}\) in the magnetic branch and \(m_{\E}\) in the electric branch. The total variation of Eq.~\eqref{eq:potential} on \(s=m\) reduces to
\begin{equation}
\left.\delta V\right|_{s=m}
=\left.\frac{\partial V}{\partial s_{ij}}\right|_{s=m}\delta s_{ij}
  +\det m\,\tr(m^{-1}\delta m)
=\delta\det m .
\label{eq:cw-auxiliary}
\end{equation}
Stationarity removes every \(\delta s\) term, including those proportional to derivatives of \(\chi\). The time-constraint obstruction is therefore the same with the auxiliary tensor retained or eliminated.

\subsection{Exact global dilations with a transforming clock}
\label{sec:cw-global}

Consider a constant positive scaling factor \(\lambda\), take \(X^I\) of weight \(1\), and assign weights \(w_t,w_A\) to \(t,A_i\). The momenta have the inverse weights. At fixed tension the clock square, electric flux square and electric determinant scale as \(\lambda^{-2w_t}\), \(\lambda^{2-2w_A}\) and \(\lambda^{2p}\). For the magnetic determinant, a term with \(k\) magnetic two-planes scales as \(\lambda^{2(p-2k)+2w_Ak}\). The zero-flux term fixes \(w_t=-p\). For generic magnetic fields the \(k=1\) term then fixes \(w_{A,\M}=2\). In the electric action the flux term fixes \(w_{A,\E}=1-p\). The two dilations are therefore
\begin{equation}
\begin{gathered}
X^I\mapsto\lambda X^I,\qquad t\mapsto\lambda^{-p}t,\qquad
A_a\mapsto\lambda^{w_A} A_a,\qquad
w_{A,\M}=2,\quad w_{A,\E}=1-p,\\
s\mapsto\lambda^2s,\qquad e\mapsto\lambda^{-2p}e,\qquad N^i\mapsto N^i .
\end{gathered}
\label{eq:cw-global-weights}
\end{equation}
Both spatial blocks and \(s\) have weight two, so \(s^{-1}m\) is invariant and \(V\) has weight \(2p\) throughout the positive-determinant domain. The clock square and, in the electric branch, the flux square have the same weight. The lapse compensates it. The canonical kinetic terms and spatial constraint have weight zero, and the weight of \(A_0\) cancels that of Gauss. The auxiliary action and auxiliary constraint surface are therefore invariant under the constant transformation. On a domain with invariant boundary conditions the conserved charge is
\begin{equation}
Q_{w_t,w_A}=\int_\Sigma(w_tP_t t+P_I X^I+w_A\pi^i A_i).
\label{eq:cw-weighted-charge}
\end{equation}
Extending the generator to act on the auxiliary canonical pair adds \(2\int_\Sigma\rho^{ij}s_{ij}\), which vanishes on the primary surface. Under \(A_i\mapsto A_i+\partial_i\alpha\), integration by parts gives the charge variation \(w_A\int_\Sigma\pi^i\partial_i\alpha=w_A\mathcal G[\alpha]\), with the sign of Eq.~\eqref{eq:generators}. Thus the charge is invariant under small gauge transformations on the Gauss surface. A fixed nonzero quantized magnetic flux can restrict the global scaling, since \(\int F\) over a closed two-cycle scales by \(\lambda^{w_A}\).

Promoting the dilation parameter to a function gives \(-w_A(\chi\mathcal G+\pi^i\partial_i\chi)\) in the Gauss variation. The spatial variation follows from Eq.~\eqref{eq:cw-gauss-spatial} after replacing \(P_I X^I+\pi^jA_j\) by \(w_tP_tt+P_I X^I+w_A\pi^jA_j\) and multiplying the last term by \(w_A\). Both branches have \(w_A\ne0\) for \(p>1\). On the planar constraint surface, a constant flux along the gradient of \(\chi\) makes the Gauss variation nonzero. A parameter constant in space but varying in time produces the kinetic variation \(\dot\chi Q_{w_t,w_A}\). Shifting the constant clock on the planar data changes the charge density by \(w_tP_t\,\delta t\) while preserving the constraints. The kinetic variation thus obstructs time localization as well. The constant dilations are global symmetries on compatible domains.

The dilation weights also distinguish the limit from the finite-parameter actions in Eqs.~\eqref{eq:cmc} and \eqref{eq:cec}. At fixed \(c>0\), the term \(c^2P_IP_I\) carries weight \(-2\), and the clock-gradient term in \(G_c\) carries weight \(-2p\), different from the weights \(2p\) and \(2\) of their respective sums. If one assigns \(c\) weight \(p+1\), then both terms acquire the required weights. In the electric branch the combination \(cF_{\spat}\) also acquires weight \(2\). In the magnetic branch the factor \(c^2\pi^iG_{c,ij}\pi^j\) acquires weight \(2(p+1)-4+2=2p\). This is a covariance between actions with different parameter values. At \(c=0\) the limiting action has the constant dilation at fixed contraction parameter.

The dilations extend to both flat Type II actions in Eqs.~\eqref{eq:super-action} and \eqref{eq:super-electric-action}. Give \(\Theta\) weight \(-p/2\) and its momentum the opposite weight. In the magnetic action each spatial Clifford one-form has weight one and \(F\) has weight two, so \(\boldsymbol K\) has weight \(p\). In the electric action \(\boldsymbol K_0\) has weight \(p\), and \(\boldsymbol L_\pi\) has weight \(p\) from the weight \(p-1\) of \(\pi^i\) and the weight one of \(\partial_iX^I\). Both fermion kinetic matrices therefore have the weight \(p\) of \(P_t\), which cancels the weight of the two spinors. The odd constraints have weight \(p/2\) and the normalized \(\kappa\) projectors are invariant. A regular parity-preserving scaling leaves the bosonic slice invariant in the absence of external odd data. Its bosonic constraint variation therefore still obstructs a local extension of Eq.~\eqref{eq:cw-charge}.

\section{Conclusion}

The two Carroll contractions of the Polyakov action have the same auxiliary-constraint rank on the regular domain at fixed positive rescaled tension. In each branch the order of contraction and elimination leaves the matter symplectic form and gauge generators unchanged. The proof therefore establishes equivalence on the local reduced phase space.

In target-time gauge the spatial embeddings are time independent in both branches. The magnetic determinant drives the gauge momentum at fixed field strength. The electric displacement is constant in time and drives the gauge potential. On a common static warped background, the magnetic gauge and embedding forces jointly preserve the tangential constraint. The flux-aligned electric solution feels the normal gradient of its effective line tension.

Both Type II theories retain sixteen independent \(\kappa\)-symmetry parameters. Their fermionic constraints and supersymmetry algebras reflect the gauge scaling. The magnetic matrices contain the full spatial RR polynomial, and the electric matrices contain the zero-field RR term and the NS flux term. At fixed circle radius, spatial T-duality identifies the reduced symplectic forms, constraints and charges on circle zero modes. The magnetic branch retains the gradient of the dual scalar, and the electric branch retains its momentum after the corresponding rescaling. The prescribed local Carroll--Weyl transformation is obstructed in both branches, while the constant dilations extend to both supersymmetric actions.

\acknowledgments
Limin Zeng would like to express his deepest and foremost thanks to his
partner Lu Lu.

\appendix
\renewcommand{\theequation}{\Alph{section}.\arabic{equation}}
\section{Direct DBI momentum check}
\label{app:direct}

We verify the reduced finite-parameter constraint directly from the DBI density. Let \(\Delta_{\rm DBI}=\sqrt{-\det M}\), with time column \(a^b=(M^{-1})^{b0}\) and transposed time row \(b^a=(M^{-1})^{0a}\). Write \(S=(a+b)/2\) and \(L=(a-b)/2\). The determinant derivative contracts the embedding velocities with \(S\) and the gauge velocities with \(L\), giving
\begin{equation}
P_\mu=-T\Delta_{\rm DBI} S^a\partial_a X_\mu,\qquad
\pi^i=-T\Delta_{\rm DBI} L^i,\qquad L^0=0.
\end{equation}
Adding and subtracting \(Ma=\mathbf e_0\) and \(M^{\mathrm{T}}b=\mathbf e_0\) isolates the symmetric and antisymmetric combinations.
\begin{equation}
GS+FL=\mathbf e_0,\qquad GL+FS=0 .
\end{equation}
Contract the first equation with \(S\) and the second with \(L\). The mixed terms cancel by antisymmetry of \(F\). The \(00\) entry of \(M^{-1}\) is the ratio of the corresponding cofactor of \(M\) to its determinant.
\begin{equation}
S^{\mathrm{T}}GS+L^{\mathrm{T}}GL=S^0=\frac{\det m_{\rm rel}}{\det M}.
\end{equation}
It follows that \(P_\mu g^{\mu\nu}P_\nu+\pi^iG_{ij}\pi^j=-T^2\det m_{\rm rel}\). The spatial components of the first inverse identity give the spatial constraint. Independently, \(A_0\) enters through \(F_{0i}\), so its variation is \(-\int\dd^{p+1}\sigma\,\pi^i\partial_i\delta A_0\). Integrating by parts gives \(\partial_i\pi^i=0\). This recovers the time, spatial and Gauss constraints of Eqs.~\eqref{eq:finiteC} and \eqref{eq:spatial-gauss}.

\section{Limiting rank of the original auxiliary tensor}
\label{app:blocks}

The block inverse of Eq.~\eqref{eq:schur}, multiplied by \((-\det Q)^{1/(p-1)}\), reconstructs \(k\). With the common clock, lapse and tension scalings this gives
\begin{equation}
k=\begin{pmatrix}
r^{\mathrm{T}}sl-(cT_0 e)^2\det s&-r^{\mathrm{T}}s\\
-sl&s
\end{pmatrix},\qquad l=-N+W,\qquad r=-N-W .
\end{equation}
Since \(W=c^2w\) in the magnetic branch and \(W=cw\) in the electric branch, the limit at fixed \(N,s\) is the same
\begin{equation}
k_0=\begin{pmatrix}
N^{\mathrm{T}}sN&N^{\mathrm{T}}s\\
sN&s
\end{pmatrix}.
\end{equation}
The vector with time component one and spatial part \(-N\) annihilates \(k_0\) from both sides. Since the spatial block is invertible, \(k_0\) has rank \(p\). The rescaled variables \(s,w,e,N\) stay finite. Sections~\ref{sec:magnetic} and \ref{sec:electric} give their first-order kinetic and gauge terms.

\section{Functional derivatives and the bosonic contraction}
\label{app:bosonic-time-details}

Use the unscaled parent variables for the functional derivatives below. Write \(X_i^\mu=\partial_iX^\mu\) and \(L_j=G_{jk}\pi^k\). In the flat target,
\begin{equation}
\delta G_{ij}=X_{i\mu}\partial_j\delta X^\mu+X_{j\mu}\partial_i\delta X^\mu,
\qquad
\delta F_{ij}=\partial_i\delta A_j-\partial_j\delta A_i .
\end{equation}
Vary the electric term as
\(\delta(\pi^iG_{ij}\pi^j)/2=L_j\delta\pi^j+\pi^i\pi^j\delta G_{ij}/2\).
In the determinant variation, the two metric contributions give equal contractions with the symmetric part of \(R\). The two field-strength contributions give equal contractions with its antisymmetric part. Each factor of two cancels the one half in the time generator. In terms of the currents of Eq.~\eqref{eq:finite-gradient-currents}, the full variation is
\begin{equation}
\delta H_{\rm rel}[f]=\int_\Sigma f
\left(P^\mu\delta P_\mu+L_j\delta\pi^j+
J_\mu^i\partial_i\delta X^\mu+K^{ij}\partial_i\delta A_j\right).
\end{equation}
For a closed slice or compact supports, integrating the last two terms by parts and inserting all four derivatives in the canonical bracket gives
\begin{equation}
\begin{aligned}
\{H_{\rm rel}[f],H_{\rm rel}[g]\}
={}&\int_\Sigma P^\mu\bigl[f\partial_i(gJ_\mu^i)-g\partial_i(fJ_\mu^i)\bigr]\\
&+\int_\Sigma L_j\bigl[f\partial_i(gK^{ij})-g\partial_i(fK^{ij})\bigr].
\end{aligned}
\end{equation}
For either current \(Y\), the product rule separates the bracketed expression into \(b_iY+(fg-gf)\partial_iY\), with \(b_i=f\partial_i g-g\partial_i f\). The second term vanishes, including when the derivative acts on the inverse matrix in the current. The inverse identities give
\begin{equation}
(R-R^{\mathrm{T}})G_{\spat}=(R+R^{\mathrm{T}})F_{\spat},
\qquad
K^{ij}L_j=T^2d_{\rm rel} R^{(ij)}F_{jk}\pi^k=\mathcal S^{ij}F_{jk}\pi^k.
\end{equation}
The final equality uses \(\pi^jF_{jk}\pi^k=0\). The embedding term is \(J_\mu^iP^\mu=\mathcal S^{ij}P_\mu X_j^\mu\), so both terms combine into Eq.~\eqref{eq:finite-time-algebra} without imposing any matter constraint.

The clock and gauge rescalings preserve both terms \(P_0^{\rm rel}\delta X^0_{\rm rel}\) and \(\pi^i_{\rm rel}\delta A_i^{\rm rel}\) in the symplectic potential. They also preserve \(P_0^{\rm rel}\partial_i X^0_{\rm rel}\) and \(F^{\rm rel}_{ij}\pi^j_{\rm rel}\) in the spatial density. The two time generators contribute \(c^4\) to the structure function in the bracket. Substituting \(T_{\rm rel}=T_0/c\) gives Eq.~\eqref{eq:contracted-time-structure}. The product \(\det(m_{\alpha_{\rm g},c})\operatorname{sym}(m_{\alpha_{\rm g},c}^{-1})\) is the symmetric adjugate and is polynomial in the entries of \(m_{\alpha_{\rm g},c}\). The leading coefficients are
\begin{equation}
\begin{aligned}
\lim_{c\to0}c^{-2}\mathcal S_{\M,c}^{ij}
&=T_0^2\det(h+F_{\spat})\bigl[(h+F_{\spat})^{-1}\bigr]^{(ij)},\\
\lim_{c\to0}c^{-2}\mathcal S_{\E,c}^{ij}
&=\pi^i\pi^j+T_0^2\det h\,(h^{-1})^{ij}.
\end{aligned}
\end{equation}
The finite bracket is therefore quadratically suppressed for bounded smooth data and fixed smearings. The limiting constraints commute strongly by the canonical-variable argument for both branches in Sec.~\ref{sec:reduction}.

\section{Superbrane symplectic and Dirac calculation}
\label{app:super-time-details}

\subsection{Configuration terms in the symplectic form}

The symbol \(\delta\) denotes the field-space exterior derivative here. Use even variations. A fermionic constraint smeared by an odd external spinor also generates an even flow. Let \(V_Z\) be the superspace component of a configuration variation and \(Z_i=\partial_i Z\). Use \(\Omega=\delta\vartheta\) with \(\iota_{X_{F_1}}\Omega=-\delta F_1\) for a Hamiltonian functional \(F_1\). The variation of Eq.~\eqref{eq:super-horizontal-theta} splits into terms containing momentum variations and a configuration curvature.
\begin{equation}
\begin{aligned}
\Omega={}&\int_\Sigma
\left(\delta\widehat P_M\wedge\Pi_\delta^M+
\delta\widehat E^i\wedge\mathfrak a_i+
\delta(\Psi_{\rm rel}\,\delta\theta)\right)+\mathcal B,\\
\mathcal B(V,W)={}&\int_\Sigma
\left[\widehat P_M(\dd\Pi^M)(V_Z,W_Z)+
\widehat E^i(\delta\mathfrak a_i)(V,W)\right]
+\delta\vartheta_{\rm WZ}(V,W).
\end{aligned}
\end{equation}
In this evaluation the configuration variations hold \(\widehat P,\widehat E,\Psi_{\rm rel}\) fixed. With \(H_3=\dd B_2\), the pullback exterior derivative gives
\begin{equation}
\begin{aligned}
(\delta\mathfrak a_i)(V,W)
&=-H_3(V_Z,W_Z,Z_i)+\partial_i[B_2(V_Z,W_Z)],\\
\int_\Sigma\widehat E^i(\delta\mathfrak a_i)(V,W)
&=-\int_\Sigma\left[\widehat E^iH_3(V_Z,W_Z,Z_i)
+(\partial_i\widehat E^i)B_2(V_Z,W_Z)\right].
\end{aligned}
\end{equation}
Spatial integration by parts moves the divergence in the first line onto the mechanical flux.

For the Wess--Zumino term, take the product of the configuration family and the spatial slice. Write \(\mathbf d\) for the total exterior derivative, with field differentials ordered first, and extend the spatial potential as \(\mathbf A=A_i\dd\sigma^i\). Use the RR polyform coupling, in which a sum of form potentials couples through an exterior exponential of the field strength \cite{GreenHullTownsend1996}, to define
\begin{equation}
\mathcal F_{\rm tot}=\mathbf d\mathbf A-Z^*B_2,\qquad
\mathbf d\mathcal F_{\rm tot}=-Z^*H_3,\qquad
\mathcal W=T[Z^*C\wedge e^{\mathcal F_{\rm tot}}]_{p+1},
\end{equation}
Here \(C\) denotes the local RR potentials, and \((r,s)\) counts field and spatial differentials. The spatial integral of the one-variation component gives \(\vartheta_{\rm WZ}\). The two-variation component of the total derivative is
\begin{equation}
(\mathbf d\mathcal W)_{2,p}
=\delta\mathcal W_{1,p}+\dd_\Sigma\mathcal W_{2,p-1},
\qquad
\delta\vartheta_{\rm WZ}=\int_\Sigma(\mathbf d\mathcal W)_{2,p}.
\end{equation}
The second identity uses the slice boundary condition. The mixed and two-variation components of the total field strength are
\begin{equation}
\mathcal F_{\rm tot}(V,\partial_i)=\mathfrak a_i(V),\qquad
\mathcal F_{\rm tot}(V,W)=-B_2(V_Z,W_Z).
\end{equation}
The exponential contributes \(-Z^*H_3\wedge e^{\mathcal F_{\rm tot}}\) upon differentiation, using \(\mathbf d\mathcal F_{\rm tot}=-Z^*H_3\). The derivative of the RR potential contributes \(Z^*(\dd C)\). The components of \(\mathcal F_{\rm tot}\) with at least one field-space index contain only \(\delta A_i\) and contractions of \(B_2\). Thus the integrated curvature \(\mathcal B(V,W)\) contains no derivatives of the variations.

\subsection{Bracket of the two time generators}

For two even Hamiltonian functionals \(F_1,F_2\), let \(V_{F_1},V_{F_2}\) be the configuration projections of their Hamiltonian vector fields. The derivative \(\delta_V^{\rm conf}\) acts at fixed \((\widehat P,\widehat E,\Psi_{\rm rel})\). The canonical momentum-variation terms and \(\iota_{X_{F_1}}\Omega=-\delta F_1\) give
\begin{equation}
\{F_1,F_2\}=\delta^{\rm conf}_{V_{F_2}}F_1-\delta^{\rm conf}_{V_{F_1}}F_2
+\mathcal B(V_{F_1},V_{F_2}).
\label{eq:horizontal-bracket-rule}
\end{equation}
Testing the symplectic equation on pure momentum variations gives \(V_{F_1}\) from the momentum derivatives of \(F_1\). In \(\delta F_1(X_{F_2})\), the momentum component of \(X_{F_2}\) then pairs with \(V_{F_1}\). Testing the equation for \(X_{F_2}\) on a fixed-momentum lift of \(V_{F_1}\) gives \(-\delta^{\rm conf}_{V_{F_1}}F_2+\mathcal B(V_{F_1},V_{F_2})\) for that pairing. This establishes Eq.~\eqref{eq:horizontal-bracket-rule} for the even generators, including the spinor-smeared constraints. The time generators have configuration projections \(V_f=fV_1\) and \(V_g=gV_1\), where \(V_1\) is the configuration projection for unit smearing, by the momentum derivatives given after Eq.~\eqref{eq:super-horizontal-theta}. The local antisymmetric curvature therefore satisfies \(\mathcal B(V_f,V_g)=0\). At fixed momenta the Hamiltonian variation is
\begin{equation}
\begin{gathered}
\delta^{\rm conf}_V H_{\rm rel}^{\rm susy}[f]
=\int_\Sigma f\left[J_M^i\delta_V\Pi_i^M+\tfrac12 K^{ij}\delta_V\mathcal F_{ij}\right],\\
J_M^i=\mathcal S_{\rm susy}^{ij}\Pi_{jM},\qquad K^{ij}=T^2\det m_{\rm susy}\,R^{[ij]} .
\end{gathered}
\end{equation}
Cartan's formula for the variation of a pullback gives
\begin{equation}
\begin{aligned}
\delta_V\Pi_i^M&=\partial_i[\Pi_\delta^M(V)]+(\dd\Pi^M)(V_Z,Z_i),\\
\delta_V\mathcal F_{ij}
&=\partial_i[\mathfrak a_j(V)]-\partial_j[\mathfrak a_i(V)]-H_3(V_Z,Z_i,Z_j).
\end{aligned}
\end{equation}
Putting \(L_i=G_{ij}\widehat E^j\), the two variations for \(V_g\) are
\begin{equation}
\begin{aligned}
\delta_g\Pi_i^M&=(\partial_i g)\widehat P^M
   +g\left[\partial_i\widehat P^M+(\dd\Pi^M)(V_{1,Z},Z_i)\right],\\
\delta_g\mathcal F_{ij}&=(\partial_i g)L_j-(\partial_j g)L_i
   +g\left[\partial_i L_j-\partial_j L_i-H_3(V_{1,Z},Z_i,Z_j)\right].
\end{aligned}
\end{equation}
All terms in square brackets cancel between the two fixed-momentum derivatives in Eq.~\eqref{eq:horizontal-bracket-rule}. Antisymmetry of \(K\) combines the two differentiated gauge terms, leaving \(b_i(J_M^i\widehat P^M+K^{ij}L_j)\). The inverse identity in Appendix~\ref{app:bosonic-time-details} and \(\widehat E^i\mathcal F_{ij}\widehat E^j=0\) turn this into the mechanical spatial generator in Eq.~\eqref{eq:finite-super-time-bracket}. For the spatial generator, take \(V_Z=v^iZ_i\) and \(V(A_j)=v^i F_{ij}\). Then \(\mathfrak a_j(V)=v^i\mathcal F_{ij}\), and the Wess--Zumino evaluation has a repeated spatial tangent and vanishes. Evaluating the remaining terms of the one-form gives \(C_{\rm mech}[v]+\int_\Sigma v^i\Psi_{\rm rel}\partial_i\theta\).

\subsection{Fermionic bracket and its rank near both limits}

For two odd, field-independent spinor smearings, both configuration derivatives of \(\Psi_{\rm rel}[\lambda]\) and \(\Psi_{\rm rel}[\eta]\) at fixed momenta vanish. Equation~\eqref{eq:horizontal-bracket-rule} then gives the exact smeared expression
\begin{equation}
\{\Psi_{\rm rel}[\lambda],\Psi_{\rm rel}[\eta]\}
=\mathcal B_\Pi(V_\lambda,V_\eta)
+\mathcal B_{\rm NS}(V_\lambda,V_\eta)
+\int_\Sigma(\mathbf d\mathcal W)_{2,p}(V_\lambda,V_\eta),
\end{equation}
Here \(\mathcal B_\Pi\) and \(\mathcal B_{\rm NS}\) are the supervielbein and NS contributions above. Each flow depends algebraically on its spinor smearing, so this expression gives a local fermionic bracket. The Clifford kernel and stabilization identities are given in the Type IIA analysis \cite{HatsudaKamimura1998} and the Type IIB analysis \cite{KamimuraHatsuda1998}. On the Gauss surface their kernel is \(2{\bf C}\Xi\,\delta_\Sigma\), with
\begin{equation}
\Xi=\widehat P_M\Gamma^M+\Lambda\widehat E^i\Pi_i^M\Gamma_M
        +T[\mathcal U_{p,\mathrm{rel}}]_{\spat}.
\end{equation}
Here \(\Lambda=\Gamma_{11}\) in IIA and \(\Lambda=\tau_3\) in IIB. The last term uses the spatial Clifford polynomial of Eq.~\eqref{eq:super-polynomial} with exact superforms in place of their bosonic limits. The additional term off the Gauss surface is proportional to the divergence of the total electric flux and quadratic in the spinors. In IIA the product with the companion matrix \(\widetilde\Xi\) is a linear combination of the time and spatial constraints, with a fermionic improvement of the time constraint in the canonical formula. In IIB the corresponding product is \(\Xi^2\). On the full constraint surface the companion or nilpotent matrix supplies the kappa null directions. The time improvement in both formulations has the form \(\Psi_{\rm rel}\,\mathcal T^i\partial_i\theta\), where \(\mathcal T^i\) is the matrix from the finite stabilization equation.

Under \(\theta_{\rm rel}=\sqrt c\,\Theta\), the row momentum is \(\zeta=\sqrt c\,\zeta_{\rm rel}\), with \(\zeta_{\rm rel}\) conjugate to \(\theta_{\rm rel}\). The finite rescaled constraint is \(\Psi_c=\sqrt c\,\Psi_{\rm rel}\), so its bracket carries one factor of \(c\). In the Clifford kernel the mechanical time term becomes \(P_t^{\rm mech}\one\). Bounded spatial mechanical momenta contribute \(O(c)\). In the magnetic scaling the flux term is \(O(c)\) as well, and the Wess--Zumino term tends to \(T_0\boldsymbol K\). In the electric scaling the flux contraction tends to \({\bf C}\Lambda\Gamma_I\pi^i\partial_iX^I\), because \(c\widehat E^i\to\pi^i\) and the time vielbein is \(O(c)\). Only the term with no spatial super field-strength factors contributes to the RR limit \(T_0\boldsymbol K_0\). The Gauss term is bilinear in the parent spinors and proportional to \(\partial_i\pi_{\rm rel}^i\). With bounded rescaled flux divergence it is \(O(c^2)\) in the magnetic scaling and \(O(c)\) in the electric scaling. Both limits have \(P_t^{\rm mech}\to P_t\). The kernels are thus
\begin{equation}
\lim_{c\to0}\{\Psi_c,\Psi_c\}
=\begin{cases}
2(P_t\one+T_0\boldsymbol K)\delta_\Sigma & \text{magnetic},\\
2(P_t\one+T_0\boldsymbol K_0+\boldsymbol L_\pi)\delta_\Sigma & \text{electric}.
\end{cases}
\end{equation}

On the future branch the magnetic and electric minus blocks are \(-4T_0\Delta\,\boldsymbol P_{\M,-}\) and \(-4\varepsilon_{\E}\boldsymbol P_{\E,-}\), respectively. Both are invertible on their images. By continuity the finite rescaled minus block has invertible body for sufficiently small \(c\) along any smooth regular family. Its full inverse follows from a finite geometric series in the nilpotent correction. This block gives at least sixteen independent directions, while the finite \(\kappa\) directions bound the on-shell rank by sixteen.

\subsection{Second-class correction and the strong Carroll bracket}

For a field-independent odd smearing, choose \(V_\lambda(\theta)=\lambda\). The other momentum derivatives give \(\Pi_\delta(V_\lambda)=\mathfrak a_i(V_\lambda)=0\). Cartan variation yields the mixed bracket
\begin{equation}
\begin{aligned}
\{H_{\rm rel}^{\rm susy}[f],\Psi_{\rm rel}[\lambda]\}={}&\int_\Sigma f\left[J_M^i(\dd\Pi^M)(V_{\lambda,Z},Z_i)
       -\tfrac12 K^{ij}H_3(V_{\lambda,Z},Z_i,Z_j)\right]\\
&+\mathcal B(V_f,V_\lambda).
\end{aligned}
\end{equation}
At fixed residual momentum, \(\Psi_{\rm rel}[\lambda]\) has zero configuration derivative. Neither smearing is differentiated. Projection adds terms proportional to \(\Psi_{\rm rel}\), so \(\{\Psi_{{\rm rel},\alpha}^{-},H_{\rm rel}^{\rm susy}[f]\}\simeq fM_\alpha\). The symmetric Type II kernel \(C_{-\alpha\beta}(x)\delta_\Sigma(x,y)\) has a local symmetric inverse in the regular region. The Dirac correction is
\begin{equation}
\begin{multlined}
-\int_{\Sigma^2}\{H_{\rm rel}^{\rm susy}[f],\Psi_{{\rm rel},\alpha}^{-}(x)\}
 (C_-^{-1})^{\alpha\beta}(x,y)\\
\quad\times\{\Psi_{{\rm rel},\beta}^{-}(y),H_{\rm rel}^{\rm susy}[g]\}
\\
\simeq \int_\Sigma fg\,M_\alpha(C_-^{-1})^{\alpha\beta}M_\beta
=0 .
\end{multlined}
\end{equation}
The odd coefficients anticommute, so their contraction with the even symmetric inverse vanishes. The first-class bracket thus closes weakly on the canonical spatial generator.

The polynomial cofactor in Eq.~\eqref{eq:super-time-structure} stays finite at each spinor order, so the bracket vanishes in both limits.

\bibliographystyle{JHEP}
\bibliography{direct_carroll_dbrane_en}

\end{document}